\documentclass[twocolumn,twocolappendix,trackchanges]{aastex701}
\usepackage[fleqn]{amsmath}
\usepackage{amssymb,bm}
\usepackage{bmpsize}
\usepackage{graphicx}
\usepackage{physics}
\usepackage{siunitx}
\usepackage{bookmark}
\usepackage{caption}
\usepackage{subcaption}
\usepackage{booktabs}
\usepackage{amsmath}

\newcommand{\Sigmag}{\Sigma_{\mathrm{g}}}

\newcommand{\rhog}{\rho_{\mathrm{g}}}
\newcommand{\cs}{c_{\mathrm{s}}}
\newcommand{\Om}{\Omega}

\newcommand{\Hd}{H_{\mathrm{d}}}
\newcommand{\etao}{\eta_{\mathrm{O}}}
\newcommand{\etaa}{\eta_{\mathrm{A}}}
\newcommand{\baretao}{\bar{\eta}_{\mathrm{O}}}
\newcommand{\baretaa}{\bar{\eta}_{\mathrm{A}}}

\newcommand{\St}{\mathrm{St}}

\begin{document}

\title{Dependence of Dust Growth on Magnetic field Strength in Protoplanetary Disks}

\author[orcid=0009-0003-0170-9919]{Takashi Shimokawa}
\affiliation{Department of Earth and Planetary Sciences, Graduate School of Sciences, Kyushu University, Fukuoka 819-0395, Japan}
\email{shimokawa.takashi.063@s.kyushu-u.ac.jp}

\author[orcid=0000-0002-9424-1192]{Yoshihiro Kawasaki}
\affiliation{Department of Earth and Planetary Sciences, Graduate School of Sciences, Kyushu University, Fukuoka 819-0395, Japan}
\email{shimokawa.takashi.063@s.kyushu-u.ac.jp}

\author[orcid=0000-0002-0963-0872]{Masahiro N. Machida}
\affiliation{Department of Earth and Planetary Sciences, Faculty of Sciences, Kyushu University, Fukuoka 819-0395, Japan}
\email{masahiro.machida.018@m.kyushu-u.ac.jp}

\begin{abstract}

We performed one-dimensional calculations to investigate how the magnetorotational instability (MRI) affects dust growth in protoplanetary disks. We simultaneously solved the evolution of the dust size distribution and the vertical magnetic field. The dust size distribution is evolved taking into account coagulation and fragmentation. For the magnetic field evolution, we considered both ambipolar diffusion and Ohmic dissipation. The MRI-driven turbulent viscosity was estimated self-consistently from the local magnetic field strength and non-ideal magnetohydrodynamics diffusivities. We found that MRI-driven turbulence becomes stronger for smaller initial plasma beta, and the enhanced turbulence promotes dust growth in MRI-active regions. However, dust growth is ultimately limited by radial drift, and the dependence of dust size on the magnetic field strength is significant only in the outer regions of the disk. In other words, over most of the disk, radial drift remains the dominant factor controlling the maximum dust size, and the magnetic field strength plays only a minor role in determining the overall dust size distribution. 

\end{abstract}

\section{Introduction} 
Planets are formed in protoplanetary disks, where submicron-sized dust grains are thought to grow through coagulation into kilometer-sized planetesimals. Understanding the time evolution of the dust size distribution is therefore a key issue in planet formation theory.

However, dust growth is not fully understood; dust grains must overcome several barriers during their evolution. In particular, two major barriers have been extensively studied. The first is the radial drift barrier, in which dust grains migrate toward the central star faster than they grow \citep{Adachi1976, Weidenschilling1977}. The second is the fragmentation barrier, in which high-velocity collisions between dust grains lead to disruption rather than coagulation \citep{Brauer2008, Wada2009, Wada2013, Hasegawa2021, Hasegawa2023}. Understanding how dust grains evolve in the presence of these barriers is therefore essential for characterizing the dust size distribution in protoplanetary disks.

The relative velocities between dust grains, which determine whether collisions result in coagulation or fragmentation, are in part controlled by turbulence in the disk. Among the mechanisms proposed to drive turbulence, magnetorotational instability (MRI ;  \citealt{BalbusHawley1991}) is considered one of the most important processes.
However, the gas in protoplanetary disks is weakly ionized, and non-ideal magnetohydrodynamic(MHD) effects such as Ohmic resistivity and ambipolar diffusion can suppress MRI activity. In the midplane of the inner disk, Ohmic resistivity dominates and suppresses MRI activity, leading to the formation of a magnetically inactive region known as a dead zone \citep{Gammie1996, Sano2000}. In the outer disk, where ambipolar diffusion dominates, MRI activity is also suppressed \citep{Bai2013, Simon2013}. As a result, the regions where MRI can operate effectively are spatially limited.

The dust size distribution, in turn, can affect the ionization degree and thereby modify non-ideal MHD effects and MRI activity. Dust grains capture charged particles (ions and electrons) onto their surfaces, thereby reducing the ionization degree of the gas \citep{Okuzumi2009, Marchand2016, Tsukamoto2022}. In particular, coagulation depletes and fragmentation replenishes the abundance of small grains, which most effectively remove ions and electrons from the gas phase and thus strongly influence the local ionization degree \citep{Kawasaki2022,Kawasaki2023}.
MRI-driven turbulence, in turn, increases the relative velocities between dust grains. The enhanced relative velocities increase collision frequencies and can thereby accelerate dust growth \citep{Sato2016}. However, when collision velocities become sufficiently high, fragmentation becomes efficient and limits the maximum dust size \citep{OkuzumiHirose2012}.
Thus, dust growth modifies MRI activity through its effect on the ionization degree, while MRI-driven turbulence affects dust growth through grain collisions, forming a feedback loop between dust evolution and magnetic activity.

To investigate dust growth with these effects, a number of multidimensional MHD simulations incorporating dust dynamics have been conducted in recent years \citep[e.g.,][]{Lebreuilly2020, Hu2019, Marchand2023, Tsukamoto2023}.
These studies have shown that the spatial distribution of dust grains can significantly differ from that of gas \citep{Lebreuilly2020}. Because the local ionization degree depends on the local dust
abundance, this dust--gas decoupling can, in turn, feed back into
non-ideal MHD effects. This feedback has been demonstrated more directly
by simulations showing that dust growth can regulate non-ideal MHD
effects, thereby influencing disk evolution \citep{Hu2019, Hu2021, Marchand2023, Tsukamoto2023}. However, such simulations are typically limited to timescales of at most tens of thousands of years after the disk formation \citep{Marchand2023, Tsukamoto2023}, which is insufficient to follow the long-term evolution of dust growth. 

On the other hand, turbulence in protoplanetary disks is often modeled using the $\alpha$-prescription, in which the turbulent viscosity is treated as a constant parameter \citep{ShakuraSunyaev1973}. This approach has been widely adopted in one-dimensional dust evolution studies that follow the long-term evolution of the dust size distribution (e.g., \citealt{Brauer2008}; \citealt{Birnstiel2010}, \citealt{Birnstiel2012}; \citealt{Okuzumi2012,Okuzumi2016}; \citealt{Kawasaki2025}). However, little attention has been paid to how the time evolution of the magnetic field strength affects
the dust size distribution in a self-consistent manner.

In this study, we investigate how magnetic fields influence dust growth by simultaneously calculating the evolution of the dust size distribution and the induction equation for the vertical magnetic field in a one-dimensional disk model. Our model self-consistently incorporates the mutual feedback between dust growth, non-ideal MHD effects, and MRI-driven turbulence.
 The outline of this paper is as follows. 
In Section~\ref{sec:method}, we describe the details 
of our calculations, including the disk model, the 
treatments of dust size evolution and magnetic field 
evolution, and the model for MRI-driven turbulence. 
In Section~\ref{sec:results}, we present our main 
results for models with different initial magnetic 
field strengths and fragmentation velocities. 
We follow up with an in-depth discussion of these 
results, including a timescale analysis, in 
Section~\ref{sec:discussion}. 
Finally, we summarize our main conclusions in 
Section~\ref{sec:conclusion}.

\section{Method} \label{sec:method}
We investigate the co-evolution of MRI and dust growth by simultaneously calculating the time evolution of the dust size distribution and the radial distribution of the vertical magnetic field in a one-dimensional disk model. 
We note that the gas dynamics is not solved in this model; the validity of this assumption is discussed in Section \ref{subsec:time_scale}.

We assume that the disk is in vertical hydrostatic equilibrium. Although the dust evolution is described in terms of vertically integrated surface density, we evaluate the magnetic diffusivities from the vertical distributions of the gas density, dust abundance, and ionization degree. This is because dust settling produces significant vertical variations in the ionization degree, leading to a strong vertical dependence of Ohmic resistivity and ambipolar diffusivity. This vertical dependence must be taken into account because it determines the strength of MRI-driven turbulence.

We therefore first calculate the magnetic diffusivities in the $(r,z)$ plane using the vertical distribution of dust (Section \ref{subsec: calculation of non-ideal effect}) and obtain effective diffusivities for the one-dimensional radial calculation by density-weighted vertical averaging. Using these vertically averaged diffusivities, we solve the induction equation described in Section \ref{subsec:evolution of magnetic field} to obtain the radial distribution of the vertical magnetic field. We then evaluate the MRI activity and the corresponding turbulent viscosity based on the mean vertical magnetic field strength and the vertically averaged magnetic diffusivities, following the model described in Section~\ref{subsec:model of MRI}. The resulting turbulent viscosity is used to calculate the relative velocities between dust grains, including contributions from turbulence, Brownian motion, and radial drift (Section~\ref{subsec:evolution of dust size}). These relative velocities determine the coagulation and fragmentation rates and thereby the evolution of the dust size distribution. 
Finally, we update the dust surface density by accounting for both radial drift and collisional growth and fragmentation using the equation introduced in Section~\ref{subsec:evolution of dust surface density}.

\subsection{Disk model}\label{subsec:disk model}
As a disk model, we adopt the radial profiles of the gas surface density
$\Sigma_g(r)$ and temperature $T(r)$ from
\cite{Hayashi1981}:
\begin{align}
    \Sigma_g(r) &= 1.7\times10^{3}\,\mathrm{g\,cm^{-2}}
    \left(\frac{r}{1\,\mathrm{au}}\right)^{-1.5}, \label{eq:sigmag} \\
    T(r) &= 2.8\times10^{2}\,\mathrm{K}
    \left(\frac{r}{1\,\mathrm{au}}\right)^{-0.5}, \label{eq:T}
\end{align}
where $r$ is the distance from the central star.
We assume that the disk is vertically isothermal and in
hydrostatic equilibrium. Under these assumptions, the
gas volume density distribution is given by
\begin{align}
    \rho_g(r,z)
    = \frac{\Sigma_g(r)}{\sqrt{2\pi}\,H_\mathrm{g}}
    \exp\!\left(-\frac{z^2}{2H_\mathrm{g}^2}\right),
\end{align}
where $z$ is the height above the disk midplane,
$H_\mathrm{g} = c_s/\Omega$ is the gas scale height,
$c_s = \sqrt{k_\mathrm{B}T/(\mu m_\mathrm{H})}$ is the
isothermal sound speed, and
$\Omega = \sqrt{GM_\ast/r^3}$ is the Keplerian angular
velocity. Here, $k_\mathrm{B}$ is the Boltzmann constant,
$\mu = 2.34$ is the mean molecular weight,
$m_\mathrm{H}$ is the proton mass, $G$ is the
gravitational constant, and $M_\ast = 1M_\odot$ is the
stellar mass.

\subsection{Dust model}\label{subsec:dust model}
We model dust grains as compact spheres with internal 
density $\rho_{\mathrm{in}} = 1.4\ \mathrm{g\ cm^{-3}}$, 
appropriate for H$_2$O ice, so that the mass of a grain 
with radius $a_d$ is given by 
$m = \frac{4}{3}\pi a_d^3 \rho_{\mathrm{in}}$.
In reality, dust aggregates are expected to be porous, 
which may enhance their growth efficiency compared to 
compact particles \citep{Okuzumi2012, Garcia2020, 
Michoulier2024}. We neglect porosity for simplicity 
in this study. We take the monomer radius to be $a_0 = 
1.0\times10^{-5}\,\mathrm{cm}$, which also sets the 
smallest dust size in our computational grid 
(Section~\ref{subsec:numerical settings}).

The aerodynamic coupling between a dust grain and the 
gas is characterized by the Stokes number, 
$\mathrm{St} = \Omega\,t_{\mathrm{stop}}$, where 
$t_{\mathrm{stop}}$ is the stopping time given by
\begin{align}
t_{\mathrm{stop}} =
\begin{cases}
\dfrac{a_d\rho_{\mathrm{in}}}{\rho_g\,v_{\mathrm{th}}} 
& \left(a_d < \dfrac{9}{4}\lambda_{\mathrm{mfp}}\right), 
\\[10pt]
\dfrac{4\rho_{\mathrm{in}}a_d^2}
      {9\rho_g\,\lambda_{\mathrm{mfp}}\,v_{\mathrm{th}}} 
& \left(a_d > \dfrac{9}{4}\lambda_{\mathrm{mfp}}\right),
\end{cases}
\label{eq:stoppingtime}
\end{align}
where $v_{\mathrm{th}}=\sqrt{8k_{\mathrm{B}}T/
(\pi\mu m_{\mathrm{H}})}$ is the thermal velocity of 
gas molecules, $\lambda_{\mathrm{mfp}} = 
\mu m_{\mathrm{H}}/(\sigma_{\mathrm{mol}}\rho_g)$ is 
the gas mean free path, and 
$\sigma_{\mathrm{mol}}=2.0\times10^{-15}\ \mathrm{cm^2}$ 
is the collisional cross section. The Stokes number is used throughout this study to characterize the dust drift velocity, the dust scale height, and the collision kernels (Section~\ref{subsec:evolution of dust size}).

The dust surface density with mass $m$ is denoted by
$\Sigma_{\mathrm{d}}(m)$, whose time evolution is described in Section
\ref{subsec:evolution of dust surface density}. The total dust surface
density is defined as $\Sigma_{\mathrm{d,tot}}(r,t) = \int \Sigma_{\mathrm{d}}(m)\,\mathrm{d}m$.
As the initial condition, we set
$\Sigma_{\mathrm{d,tot,ini}}(r) = f_{\mathrm{dg}}\Sigma_{g}(r)$,
where $f_{\mathrm{dg}} = 0.01$ is the initial dust-to-gas mass ratio.

\subsection{Evolution of dust surface density}
\label{subsec:evolution of dust surface density}
In this subsection, we describe the evolution of the dust surface density with the following equation,
\begin{align} 
\pdv{\Sigma_\mathrm{d}(m)}{t} 
&= -\frac{1}{r}\pdv{}{r}\left(r\,v_{r}(m)\, 
\Sigma_{\mathrm{d}}(m)\right) 
+ \left.\pdv{\Sigma_{\mathrm{d}}(m)}{t} 
\right|_{\mathrm{growth}}, \label{eq:dust_sigma}
\end{align}
where $t$ is the time, $m$ is mass of dust grain and $v_{r}(m)$ is the radial velocity of dust grains. The growth term $\left.\pdv{\Sigma_{\mathrm{d}}(m)}{t} 
\right|_{\mathrm{growth}}$ on the right-hand side arises from collisional coagulation and fragmentation of dust grains and is expressed as \citep{Smoluchowski1916, Booth2018}:
\begin{equation}
\begin{aligned}
&\left.\pdv{\Sigma_\mathrm{d}(m)}{t}\right|_{\mathrm{growth}} \\
&= \frac{1}{2}\int_{0}^{m}
m\,\bar{K}(m-m_1,m_1)
\\
&\qquad \times N(m-m_1)\,N(m_1)\,\dd m_1
\\
&- \int_{0}^{\infty} 
m\,\bar{K}(m,m_1) N(m)\,N(m_1)\,\dd m_1
\\
&+ \frac{1}{2}\int_{0}^{\infty}\int_{0}^{\infty}
m\,\bar{F}(m_1,m_2)
\\
&\qquad \times N(m_1)\,N(m_2)\,
\phi(m;m_1,m_2)\,\dd m_1\,\dd m_2
\\
&- \int_{0}^{\infty}
m\,\bar{F}(m,m_1) N(m)\,N(m_1)\,\dd m_1, \label{eq:growthterm}
\end{aligned}
\end{equation}
where $\bar{K}$ and $\bar{F}$ are the vertically integrated coagulation and fragmentation kernels, respectively, $\phi(m;m_1,m_2)$ is the fragment mass distribution produced by a collision between grains of mass $m_1$ and $m_2$, and $N(m)$ is the vertically integrated number density of dust grains with mass $m$. The derivation of $\bar{K}$ and $\bar{F}$ is described in Section \ref{subsec:evolution of dust size}.

\subsection{Evolution of dust size}\label{subsec:evolution of dust size}
In this subsection, we determine the local coagulation and
fragmentation kernels, $K_{i,j}$ and $F_{i,j}$, where the subscripts $i$ and $j$ denote dust grains of mass $m_i$ and $m_j$, respectively. The vertically integrated kernels $\bar{K}$ and $\bar{F}$, introduced in Section~\ref{subsec:evolution of dust surface density}, are then obtained from these local kernels.
These kernels represent the rates at which collisions between a grain of mass $m_i$ and one of mass $m_j$ result in coagulation and fragmentation, respectively, and depend on the relative velocity and collisional cross section of the two grains.

The kernels introduced in Section~\ref{subsec:evolution of dust
surface density} are defined as vertically integrated quantities. We
therefore first obtain the local kernels $K_{i,j}$ and $F_{i,j}$, and
then obtain the vertically integrated kernels $\bar{K}(m_i, m_j)$ and
$\bar{F}(m_i, m_j)$ following \citet{Stammler2022},
\begin{align}
\bar{K}(m_i,m_j) &= \frac{K_{i,j}}{\sqrt{2\pi
\left(H_d(m_i)^2+H_d(m_j)^2\right)}}, \label{eq:K} \\
\bar{F}(m_i,m_j) &= \frac{F_{i,j}}{\sqrt{2\pi
\left(H_d(m_i)^2+H_d(m_j)^2\right)}}. \label{eq:F}
\end{align}
where $H_d(m_i)$ and $H_d(m_j)$ are the dust scale 
heights of grains with mass $m_i$ and $m_j$, 
respectively. These are defined later in this section 
(Equation~(\ref{eq:Hd})).

Following \citet{Garaud2013}, we define the relative 
velocity between two dust grains of mass $m_i$ and $m_j$ as
\begin{align}
\Delta v_{i,j} \equiv \left| \bm{v}_i - \bm{v}_j \right| .
\label{eq:dv}
\end{align}
A collision results in coagulation when 
$\Delta v_{i,j} \le v_\mathrm{frag}$ and in 
fragmentation when $\Delta v_{i,j} > v_\mathrm{frag}$, 
where $v_\mathrm{frag}$ is the fragmentation velocity. 
We treat $v_\mathrm{frag}$ as a free parameter and 
explore two values as summarized in 
Table~\ref{tab:models}. 
The higher value 
($v_\mathrm{frag} = 30\ \mathrm{m\ s^{-1}}$) 
is appropriate for icy grains, whereas the lower 
value ($v_\mathrm{frag} = 3\ \mathrm{m\ s^{-1}}$) 
is representative of silicate grains.

The local coagulation and fragmentation kernels are then 
defined as
\begin{align}
K_{i,j} \equiv \bar{s}_{i,j}
\int_{0}^{v_\mathrm{frag}}
\Delta v_{i,j}\, P(\Delta v_{i,j})\, \mathrm{d}\Delta v_{i,j},
\label{eq:Kdef}
\\
F_{i,j} \equiv \bar{s}_{i,j}
\int_{v_\mathrm{frag}}^{\infty}
\Delta v_{i,j}\, P(\Delta v_{i,j})\, \mathrm{d}\Delta v_{i,j},
\label{eq:Fdef}
\end{align}
where $\bar{s}_{i,j} = \pi(a_i+a_j)^2$ is the collisional 
cross section of two grains with radii $a_{i}$ and $a_j$, and 
$P(\Delta v_{i,j})$ is the probability density function of 
$\Delta v_{i,j}$. Assuming that the stochastic motion of each particle is 
isotropic, \citet{Garaud2013} showed that
\begin{align}
P(\Delta v_{i,j}) &= 
\frac{1}{\sqrt{2\pi}\,\sigma_{i,j}}
\frac{\Delta v_{i,j}}{\Delta V^{D}_{i,j}}
\Bigg[
\exp\!\left( 
-\frac{\left(\Delta v_{i,j}-\Delta V^{D}_{i,j}\right)^2}
{2\sigma_{i,j}^2}
\right)
\nonumber \\
&\quad
-
\exp\!\left(
-\frac{\left(\Delta v_{i,j}+\Delta V^{D}_{i,j}\right)^2}
{2\sigma_{i,j}^2}
\right)
\Bigg],
\label{eq:pdf}
\end{align}
where $\Delta V^{D}_{i,j}$ is the drift-induced relative 
velocity, whose expression is given later in this section 
(Equation~(\ref{eq:VD})), and the dispersion $\sigma_{i,j}$ 
is given by
\begin{align}
\sigma_{i,j}^2 = \frac{\pi}{8}
\left[
\left(\Delta V^{T}_{i,j}\right)^2
+\left(\Delta V^{B}_{i,j}\right)^2
\right],
\label{eq:sigma}
\end{align}
with $\Delta V^{T}_{i,j}$ and $\Delta V^{B}_{i,j}$ the 
turbulence-induced and Brownian-motion-induced relative 
velocities, respectively, whose expressions are given later 
in this section (Equations~(\ref{eq:VT}) and (\ref{eq:VB})). 

Evaluating the integrals in Equations~(\ref{eq:Kdef}) and 
(\ref{eq:Fdef}) using Equation~(\ref{eq:pdf}) yields
\begin{align}
K_{i,j} &= \bar{s}_{i,j}\Biggl[\frac{(\Delta V^{D}_{i,j})^2
+\sigma_{i,j}^2}{2\Delta V^{D}_{i,j}}
\left\{2\,{\rm erf}\left(\frac{\Delta V^{D}_{i,j}}{\sqrt{2}\,\sigma_{i,j}}\right)-A\right\} \nonumber\\
&\quad +\sqrt{\frac{2}{\pi}}\,\sigma_{i,j}\exp\left(-\frac{(\Delta V^{D}_{i,j})^2}{2\sigma_{i,j}^2}\right)
-\frac{\sigma_{i,j}}{\sqrt{2\pi}\,\Delta V^{D}_{i,j}}B\Biggr], \label{eq:K}\\
F_{i,j}
&= \bar{s}_{i,j}
\left[
\frac{\left(\Delta V^{D}_{i,j}\right)^2+\sigma_{i,j}^2}
{2\Delta V^{D}_{i,j}}\,A
+ \frac{\sigma_{i,j}}{\sqrt{2\pi}\,\Delta V^{D}_{i,j}}
\,B
\right] ,
\end{align}
The coefficients $A$ and 
$B$ are given by
\begin{align}
A &=
\mathrm{erf}\!\left(
\frac{v_{\mathrm{frag}}+\Delta V^{D}_{i,j}}
{\sqrt{2}\,\sigma_{i,j}}
\right)
-\mathrm{erf}\!\left(
\frac{v_{\mathrm{frag}}-\Delta V^{D}_{i,j}}
{\sqrt{2}\,\sigma_{i,j}}
\right),
\\
B &=
\left(v_{\mathrm{frag}}+\Delta V^{D}_{i,j}\right)
\exp\!\left(
-\frac{\left(v_{\mathrm{frag}}
-\Delta V^{D}_{i,j}\right)^2}
{2\sigma_{i,j}^2}
\right) \nonumber \\
&\quad -
\left(v_{\mathrm{frag}}-\Delta V^{D}_{i,j}\right)
\exp\!\left(
-\frac{\left(v_{\mathrm{frag}}
+\Delta V^{D}_{i,j}\right)^2}
{2\sigma_{i,j}^2}
\right).
\end{align}

Following \citet{Wada2013}, the fraction of mass redistributed into fragments relative to that retained in the largest remnant increases with the ratio of the collision velocity to the fragmentation velocity. 
We therefore adopt the fragment mass distribution function, $\phi(m;m_i,m_j)$, introduced in Equation~(\ref{eq:growthterm}), based on the model of \citet{Wada2013} as implemented by \citet{Kawasaki2022}.

We now give the expressions for $\Delta V^{T}_{i,j}$, 
$\Delta V^{B}_{i,j}$, and $\Delta V^{D}_{i,j}$, which were 
introduced above. Following \citet{OrmelCuzzi2007}, the 
turbulence-induced relative velocity under the assumption 
of Kolmogorov turbulence is given by
\begin{align}
\Delta V^{T}_{i,j}
&= \sqrt{\left(\Delta V^{T}_{I,i,j}\right)^2
+\left(\Delta V^{T}_{II,i,j}\right)^2}, 
\label{eq:VT}\\
\left(\Delta V^{T}_{I,i,j}\right)^2
&= \bar\alpha c_s^2
\left(\frac{\mathrm{St}_i-\mathrm{St}_j}
{\mathrm{St}_i+\mathrm{St}_j}\right)
 \Bigg(
\frac{\mathrm{St}_i^2}{\mathrm{St}^{\ast}_{i,j}+\mathrm{St}_i}
-\frac{\mathrm{St}_i^2}{1+\mathrm{St}_i}
\nonumber \\
&\quad
-\frac{\mathrm{St}_j^2}{\mathrm{St}^{\ast}_{i,j}+\mathrm{St}_j}
+\frac{\mathrm{St}_j^2}{1+\mathrm{St}_j}
\Bigg),
\\
\left(\Delta V^{T}_{II,i,j}\right)^2
&= \bar\alpha c_s^2
\left(\mathrm{St}^{\ast}_{i,j}
-\mathrm{St}_\eta\right) \nonumber \\
&\quad \times \Bigg[
\frac{(\mathrm{St}^{\ast}_{i,j}+\mathrm{St}_\eta)
\mathrm{St}_i+\mathrm{St}^{\ast}_{i,j}\mathrm{St}_\eta}
{(\mathrm{St}_i+\mathrm{St}^{\ast}_{i,j})
(\mathrm{St}_i+\mathrm{St}_\eta)}
\nonumber\\
&\qquad +
\frac{(\mathrm{St}^{\ast}_{i,j}+\mathrm{St}_\eta)
\mathrm{St}_j+\mathrm{St}^{\ast}_{i,j}\mathrm{St}_\eta}
{(\mathrm{St}_j+\mathrm{St}^{\ast}_{i,j})
(\mathrm{St}_j+\mathrm{St}_\eta)}
\Bigg],
\end{align}
where $\mathrm{St}_i$ and $\mathrm{St}_j$ are the Stokes numbers of
grains with masses $m_i$ and $m_j$, respectively, and
$\bar\alpha$ is the vertically averaged turbulent viscosity parameter
defined in Section~\ref{subsec:model of MRI}.
Here, $\mathrm{St}_\eta=\Omega t_\eta$, where $t_\eta$ is the
turnover timescale of the smallest eddies, given by
\begin{align}
t_\eta=\frac{t_L}{\sqrt{\mathrm{Re}}},
\label{eq:teta}
\end{align}
with $t_L=H_g/c_s=1/\Omega$ being the turnover timescale of the
largest eddies and $\mathrm{Re}=\nu_t/\nu_m$ the Reynolds number.
The turbulent and molecular viscosities are given by
$\nu_t=\bar\alpha c_sH_g$ and
$\nu_m=\frac{1}{2}v_{\mathrm{th}}\lambda_{\mathrm{mfp}}$,
respectively, with $v_{\mathrm{th}}$ and $\lambda_{\mathrm{mfp}}$
defined in Section~\ref{subsec:dust model}.
The characteristic Stokes number $\mathrm{St}^{\ast}_{i,j}$ is given by
\begin{align}
\mathrm{St}^{\ast}_{i,j}=\max\!\left(
\mathrm{St}_\eta,
\min\left(1.6\max(\mathrm{St}_i,\mathrm{St}_j),1
\right)\right),
\end{align}
following the prescription of \citet{OrmelCuzzi2007}.

The Brownian-motion-induced relative velocity is given by
\begin{align}
\Delta V^{B}_{i,j}
= \sqrt{\frac{8k_{\mathrm{B}}T}{\pi}\left(\frac{1}{m_i}+\frac{1}{m_j}\right)} \label{eq:VB}.
\end{align}
The drift-induced relative velocity is given by
\begin{align}
\Delta V^{D}_{i,j} = \sqrt{
\left(\Delta V_{r,i,j}^{D}\right)^2
+\left(\Delta V_{\phi,i,j}^{D}\right)^2
+\left(\Delta V_{z,i,j}^{D}\right)^2}, 
\label{eq:VD}
\end{align}
where
\begin{align}
\Delta V_{r,i,j}^{D} &= v_r(m_i)-v_r(m_j),\\
\Delta V_{\phi,i,j}^{D} &= v_\phi(m_i)-v_\phi(m_j),\\
\Delta V_{z,i,j}^{D} &= v_z(m_i)-v_z(m_j),
\end{align}
and $v_r(m)$, $v_\phi(m)$, and $v_z(m)$ are the 
radial, azimuthal, and vertical drift velocities 
of a particle of mass $m$, respectively. The radial and azimuthal drift velocities, $v_r(m)$ and $v_\phi(m)$ are given by \citet{Adachi1976} and \citet{Weidenschilling1977} as
\begin{align}
v_r(m) &= -\frac{2\St}{1+\St^2}\,\eta_P r\Om \label{eq:vr},\\
v_\phi(m) &= -\frac{1}{1+\St^2}\,\eta_P r\Om \label{eq:vphi},
\end{align}
where $\eta_P$ is a dimensionless parameter that characterizes the deviation of the gas rotation velocity from the Keplerian velocity owing to the radial pressure gradient, given by
\begin{align}
\eta_P=-\frac{1}{2}\left(\frac{\cs}{r\Om}\right)^2\pdv{\ln P}{\ln r},
\end{align}
and $P$ is the gas pressure. The vertical drift velocity, $v_z(m)$, is given by \citep{Okuzumi2012}
\begin{align}
v_z(m) = -H_d(m)\,\St_{\min}\,\Omega \label{eq:vz},
\end{align}
where $\mathrm{St}_{\min} = \min(\mathrm{St}, 1/2)$, and $H_d(m)$ is the dust scale height, given by \citet{Birnstiel2010} as
\begin{align}
\Hd(m)=H_g(r)\,\min\!\left(1.0,
\sqrt{\frac{\bar\alpha(r)}{\St_{\min}(1+\St^2)}}
\right).\label{eq:Hd}
\end{align}

Ambipolar diffusion can induce relative velocities between charged dust grains and promote the removal of small grains through coagulation with larger grains in molecular cloud cores \citep{Silsbee2020,Guillet2020}. Electrostatic repulsion between charged grains can also reduce the coagulation efficiency \citep{Okuzumi2009,Okuzumi2011}. We omit the contribution of ambipolar drift to the relative velocities and the effect of the electrostatic barrier on coagulation in our dust evolution calculations. We assess these effects in Sections~\ref{subsec:effect of ambipolar_drift} and \ref{subsec:effect of electrostatic barrier}, respectively.

\subsection{Evolution of magnetic field}
\label{subsec:evolution of magnetic field}
We model the magnetic field as a mean vertical field strength $B_z(r,t)$. We assume that the magnetic field evolution is dominated by diffusion and neglect the advection term in the induction equation; the validity of this assumption is discussed in Section \ref{subsec:time_scale}. Under this assumption, the induction equation for the vertical magnetic field strength is given by
\begin{align}
\pdv{B_z(r,t)}{t}
&= \frac{1}{r}\pdv{}{r}\left(r\,
(\baretao(r,t)+\baretaa(r,t))\,\pdv{B_z(r,t)}{r} \right) \label{eq:induction},
\end{align}
where $\baretao$ and $\baretaa$ are the vertically averaged magnetic diffusivities for Ohmic resistivity and ambipolar diffusion, respectively, defined as
\begin{align}
\baretao(r,t) &= \frac{1}{\Sigmag}
\int_{-3H_g}^{3H_g}\rhog(r,z)\,\etao(r,z,t)\,\dd z, \label{eq:etaObar} \\
\baretaa(r,t) &= \frac{1}{\Sigmag}
\int_{-3H_g}^{3H_g}\rhog(r,z)\,\etaa(r,z,t)\,\dd z, \label{eq:etaAbar}
\end{align}
where $\etao(r,z,t)$ and $\etaa(r,z,t)$ are the local magnetic diffusivities for Ohmic resistivity and ambipolar diffusion, respectively. 
The integrations in Equations~(\ref{eq:etaObar}) and 
(\ref{eq:etaAbar}) are carried out over the finite range 
$-3H_g \leq z \leq 3H_g$ rather than over the formally infinite 
vertical domain. This is because $\rho_g(r,z)$ decreases as 
$\exp(-z^2/2H_g^2)$, and this range contains more than 99\% of 
the gas surface density $\Sigma_g$. The contribution from 
$|z| > 3H_g$ to $\bar{\eta}_O$ and $\bar{\eta}_A$ is therefore 
negligible.
The detailed derivation of these quantities is described in Section \ref{subsec: calculation of non-ideal effect}.

\subsection{Calculation of non-ideal magnetic diffusion coefficients}\label{subsec: calculation of non-ideal effect}
To evaluate non-ideal MHD effects, it is essential to determine the abundances of charged particles in the gas phase. Dust grains reduce the ionization degree by absorbing charged particles, thereby significantly affecting non-ideal MHD processes \citep{Okuzumi2009,Marchand2016}.

A full chemical network including charged dust is computationally expensive. In this study, we therefore assume chemical equilibrium and adopt the analytic model constructed by \citet{Tsukamoto2022} to estimate the charged particle abundances and dust charge. We then use these abundances to compute the magnetic diffusivities.

We consider X-rays, cosmic rays, and radioactive 
decay as ionization sources, such that the total 
ionization rate is given by
\begin{align}
\zeta(r,z) = \zeta_{\mathrm{XR}}(r,z)
+\zeta_{\mathrm{CR}}(r,z)
+\zeta_{\mathrm{RA}},
\label{eq:zetatot}
\end{align}
where $\zeta_{\mathrm{XR}}(r,z)$, 
$\zeta_{\mathrm{CR}}(r,z)$, and $\zeta_{\mathrm{RA}}$ 
are the ionization rates due to X-rays, cosmic rays, 
and radioactive decay, respectively. In the gas phase, H$_2$ molecules are 
the primary target of ionization.

The ionization rate due to X-rays 
is given by \citep{TurnerSano2008, Okuzumi2009}
\begin{align}
\zeta_{\mathrm{XR}}
&= \zeta_{\mathrm{XR},0}
\left(\frac{r}{1\,\mathrm{au}}\right)^{-2} 
\left(\frac{L_\mathrm{XR}}
{2.0\times10^{30}\,\mathrm{erg\,s^{-1}}}\right) 
\nonumber \\
&\times\left[
\exp\!\left(-\frac{\xi_g^{+}(r,z)}
{\chi_{\mathrm{XR}}}\right)
+\exp\!\left(-\frac{\xi_g^{-}(r,z)}
{\chi_{\mathrm{XR}}}\right)
\right] \label{eq:zetaXR},
\end{align}
where $\xi_g^{+}(r,z)$ and $\xi_g^{-}(r,z)$ are 
the gas column densities above and below the point 
$(r,z)$, respectively, given by
\begin{align}
\xi_g^{+}(r,z)&=\int_{z}^{+\infty}
\rho_g(r,z')\,\mathrm{d}z',\\
\xi_g^{-}(r,z)&=\int_{-\infty}^{z}
\rho_g(r,z')\,\mathrm{d}z'.
\end{align}
The X-ray luminosity is set to 
$L_\mathrm{XR}=2.0\times10^{30}\ 
\mathrm{erg\ s^{-1}}$, and $\chi_\mathrm{XR} = 
8.0\ \mathrm{g\ cm^{-2}}$ and 
$\zeta_{\mathrm{XR},0} = 2.6\times10^{-15}\ 
\mathrm{s^{-1}}$ are fitting parameters adopted 
from \citet{TurnerSano2008}.

The ionization rate due to cosmic rays is given by \citep{Umebayashi2009}
\begin{align}
&\zeta_{\mathrm{CR}}(r,z)
= \frac{\zeta_{\mathrm{CR},0}}{2} \nonumber \\
&\Bigg\{
\exp\!\left(-\frac{\xi_g^{+}(r,z)}{\xi_{\mathrm{CR}}}\right) 
\left[1+\left(\frac{\xi_g^{+}(r,z)}{\xi_{\mathrm{CR}}}\right)^{3/4}\right]^{-4/3}
\nonumber\\
&+
\exp\!\left(-\frac{\xi_g^{-}(r,z)}{\xi_{\mathrm{CR}}}\right)
\left[1+\left(\frac{\xi_g^{-}(r,z)}{\xi_{\mathrm{CR}}}\right)^{3/4}\right]^{-4/3}
\Bigg\} \label{eq:zetaCR},
\end{align}
where $\xi_{\mathrm{CR}}\simeq 96\,\mathrm{g\,cm^{-2}}$ and $\zeta_{\mathrm{CR},0}\simeq 1.0\times10^{-17}\,\mathrm{s^{-1}}$, respectively.

We also include ionization due to radioactive decay 
of ${}^{26}\mathrm{Al}$. Following \citet{Umebayashi2009}, 
we adopt an abundance ratio of 
${}^{26}\mathrm{Al}/{}^{27}\mathrm{Al} = 5.0\times10^{-5}$, 
which gives an ionization rate of 
$\zeta_{\mathrm{RA}} = 7\times10^{-19}\ \mathrm{s^{-1}}$.

Using the total ionization rate $\zeta(r,z)$, we calculate the number density $n_j$ and charge $Q_j$ of each charged species $j$ (including  dust grains) from the chemical equilibrium model of \citet{Tsukamoto2022}. The Ohmic, Hall, and Pedersen conductivities are then calculated from these charged particle abundances as \citep{Wardle2007}

\begin{align}
\sigma_O(r,z) &= \frac{c}{B}\sum_{j} Q_j n_j \beta_j,\\
\sigma_H(r,z) &= -\frac{c}{B}\sum_{j} \frac{Q_j n_j \beta_j^2}{1+\beta_j^2},\\
\sigma_P(r,z) &= \frac{c}{B}\sum_{j} \frac{Q_j n_j \beta_j}{1+\beta_j^2},
\end{align}
where where $c$ is the speed of light and $B$ is the magnetic field
strength. The Hall parameter $\beta_j$, defined as the ratio of the Lorentz force to the drag force due to collisions with neutrals, is given by
\begin{align}
\beta_j=\frac{Q_j B}{\gamma_{j,n}\rho_g\,m_j c},\label{eq:Hall parameter}
\end{align}
where
\begin{align}
\gamma_{j,n}=\frac{\langle\sigma v\rangle_j}{m_n+m_j},
\end{align}
$m_n$ is the mass of the neutral molecules (H$_2$), 
and $\langle\sigma v\rangle_j$ is the momentum 
transfer rate coefficient \citep{PintoGalli2008}.
Using these conductivities, and defining
\begin{align}
\sigma_\perp^2 =\sigma_H^2+\sigma_P^2,
\end{align}
Ohmic resistivity and ambipolar diffusivity are expressed as
\begin{align}
\eta_O(r,z,t) &= \frac{c^2}{4\pi \sigma_O}, \label{eq:eta_A} \\
\eta_A(r,z,t) &= \frac{c^2 \sigma_P}{4\pi \sigma_\perp^2}-\eta_O.  \label{eq:eta_O}
\end{align}

The local diffusivities $\eta_O(r,z,t)$ and $\eta_A(r,z,t)$ are then vertically averaged to obtain $\bar\eta_O(r,t)$ and $\bar\eta_A(r,t)$, as described in Section~\ref{subsec:evolution of magnetic field}. 
Substituting these into the induction equation (Equation~(\ref{eq:induction})), we obtain the radial distribution of magnetic field strength $B_z(r,t)$.

\subsection{Turbulent viscosity model driven by MRI}
\label{subsec:model of MRI}

In this subsection, we evaluate the MRI-driven 
turbulent viscosity $\alpha_\mathrm{MRI}$. We follow the model of \citet{Delage2022}, in which $\alpha_\mathrm{MRI}$ 
is expressed as a function of the local magnetic 
field strength and the non-ideal MHD diffusivities, 
Ohmic resistivity and the ambipolar diffusivity.

Because our model is one-dimensional, the quantity we ultimately require is the vertically averaged turbulent viscosity,
\begin{align}
\bar{\alpha}(r) = \frac{1}{\Sigmag}
\int_{-3H_g}^{3H_g}\rhog(r,z)\,
\alpha_{\mathrm{MRI}}(r,z)\,\dd z,
\label{eq:alpha_bar}
\end{align}
Following \citet{Delage2022}, we first summarize 
the original formulation of $\alpha_{\mathrm{MRI}}$, 
based on linear analysis \citep{SanoMiyama1999} and 
three-dimensional shearing-box simulations 
\citep[e.g.,][]{Bai2011, Mohanty2018}. 

When the MRI operates at full efficiency, the 
turbulent viscosity is well approximated as 
$\alpha_{\mathrm{MRI}} \sim 1/\beta$ 
\citep{OkuzumiHirose2011}. 
The plasma beta $\beta$, defined as the ratio of 
gas to magnetic pressure, is given by
\begin{align}
\beta = \frac{8\pi \rho_g c_s^2}{B^2}.
\end{align}
Following this approximation, \citet{Delage2022} 
expressed $\alpha_{\mathrm{MRI}}$ at an arbitrary location $(r^{\prime}, z^{\prime})$ as a step function of the plasma beta $\beta$:
\begin{align}
\alpha_{\mathrm{MRI}}(r^{\prime},z^{\prime})=
\begin{cases}
\dfrac{1}{3\beta} 
& \left(\dfrac{1}{3\beta}>\alpha_{\mathrm{hydro}}\right),
\\[6pt]
\alpha_{\mathrm{hydro}} & \text{otherwise},
\end{cases}
\label{eq:alpha_MRI_step}
\end{align}
where all quantities on the right-hand side are 
evaluated at $(r^{\prime},z^{\prime})$. 
The quantity $\alpha_{\mathrm{hydro}} = 
1.0\times10^{-4}$ is the floor value of the 
turbulent viscosity. 
This floor represents turbulence driven by other 
instabilities (e.g., the vertical shear instability; 
\citealt{Urpin1998, Lin2015}) in regions where the 
MRI is inactive.

Because Equation~(\ref{eq:alpha_MRI_step}) is a step function, its
direct use can lead to numerical instability. In this work, we
therefore replace it with the following smooth approximation:
\begin{align}
&\alpha_{\mathrm{MRI}}(r^{\prime},z^{\prime}) = 10^{x},
\label{eq:alpha_mri_smooth}\\
&x = \log_{10}\alpha_{\mathrm{hydro}}
 + \log_{10}(\alpha_{\max}/\alpha_{\mathrm{hydro}})
  \,s_1 s_2, \\[4pt]
&s_1 = 1.0 + \tanh\!
    \left\{\,0.4\,\min\!\left(
        \log_{10}\Lambda_O,\,0.0
    \right)\right\}, \\[4pt]
&s_2 = 1.0 + \tanh\!
    \left[\,0.4\,\min\!\left\{
        \log_{10}(\Lambda_{\mathrm{Am}}
        /\Lambda_{\mathrm{Am,crit}}),\,0.0
    \right\}\right],
\end{align}
where $\alpha_{\max} = 1/(3\beta)$ is the maximum turbulent viscosity at
$(r^{\prime},z^{\prime})$, attained when the MRI operates at full efficiency. The Elsasser numbers $\Lambda_O$ and $\Lambda_{\mathrm{Am}}$ evaluated at $(r^{\prime},z^{\prime})$ are defined as
\begin{align}
\Lambda_O = \frac{V_A^2}{\eta_O \Omega},
\qquad
\Lambda_{\mathrm{Am}} = \frac{V_A^2}{\eta_A \Omega},
\end{align}
where $V_A = B_z/\sqrt{4\pi\rho_g}$ is the Alfv\'{e}n velocity.
Following \citet{Delage2022}, we further define
\begin{align}
\beta_{\min}(\Lambda_{\mathrm{Am}})
= \left[
\left(\frac{50}{\Lambda_{\mathrm{Am}}^{1.2}}\right)^2
+ \left(\frac{8}{\Lambda_{\mathrm{Am}}^{0.3}} + 1\right)^2
\right]^{1/2}.
\label{eq:Am_condition}
\end{align}
We define $\Lambda_{\rm Am,crit}$ as the value of $\Lambda_{\rm Am}$
that satisfies
\begin{equation}
\beta_{\rm min}(\Lambda_{\rm Am,crit}) = \beta,
\label{eq:lambda_crit}
\end{equation}
where $\beta$ is the local plasma beta.

We now describe the physical conditions under which the approximation 
$\alpha_{\mathrm{MRI}} \approx 1/\beta$, used in 
Equation~(\ref{eq:alpha_MRI_step}), is valid. 
First, both Ohmic resistivity and ambipolar diffusivity must be sufficiently weak for the MRI to operate. Following \citet{Delage2022}, we require
\begin{align}
\Lambda_O &> 1,
\label{eq:Ohm_condition}\\
\beta &> \beta_{\min}(\Lambda_{\mathrm{Am}}).
\label{eq:beta_condition}
\end{align}
These conditions are already incorporated into the continuous approximation of Equation~(\ref{eq:alpha_mri_smooth}) through its 
dependence on $\Lambda_O$ and $\Lambda_{\mathrm{Am}}$. 
The value of $\alpha_{\mathrm{MRI}}$ evaluated from 
Equation~(\ref{eq:alpha_mri_smooth}) approaches 
$\alpha_{\mathrm{hydro}}$ when either conditions is not satisfied.

Second, for the MRI to develop, its fastest-growing wavelength must not exceed the thickness of the layer in which the MRI is active. 
We define the MRI-active layer, $\mathcal{L}_{\mathrm{MRI}}(r^{\prime})$, as the set of heights $z$ at a fixed radius $r^{\prime}$
satisfying
\begin{align}
\mathcal{L}_{\mathrm{MRI}}(r^{\prime}) =
\left\{z^{\prime} \mid
\alpha_{\mathrm{MRI}}(r^{\prime}, z^{\prime}) >
\alpha_{\mathrm{hydro}}\right\}.
\label{eq:mri_active_layer}
\end{align}
At a given radius, $\mathcal{L}_{\mathrm{MRI}}$ may consist of multiple disconnected regions.
We therefore define $h_{\mathrm{MRI}}(r^{\prime})$ as the thickness of the thickest such region within $\mathcal{L}_{\mathrm{MRI}}(r^{\prime})$. The wavelength of the fastest-growing MRI mode is given by
\begin{align}
\lambda_{\mathrm{MRI}}(r^{\prime}, z^{\prime}) = \frac{2\pi V_A}{\Omega},
\end{align}
and we denote its largest value within
$\mathcal{L}_{\mathrm{MRI}}(r^{\prime})$ by
\begin{align}
\lambda_{\mathrm{MRI,max}}(r^{\prime}) =
\max_{z^{\prime} \in \mathcal{L}_{\mathrm{MRI}}(r^{\prime})}
\left\{\lambda_{\mathrm{MRI}}(r^{\prime}, z^{\prime})\right\}.
\end{align}
The MRI is considered active at radius $r^{\prime}$ when
\begin{align}
\lambda_{\mathrm{MRI,max}}(r^{\prime})
< \min(H_g, h_{\mathrm{MRI}}(r^{\prime})).
\label{eq:MRI_condition}
\end{align}
When this condition is not satisfied, the MRI is considered inactive, and
we set $\alpha_{\mathrm{MRI}}(r^{\prime}, z^{\prime}) =
\alpha_{\mathrm{hydro}}$, regardless of the value given by Equation~(\ref{eq:alpha_mri_smooth}).

\begin{figure}[t]
  \centering
  \includegraphics[width=0.75\linewidth]
  {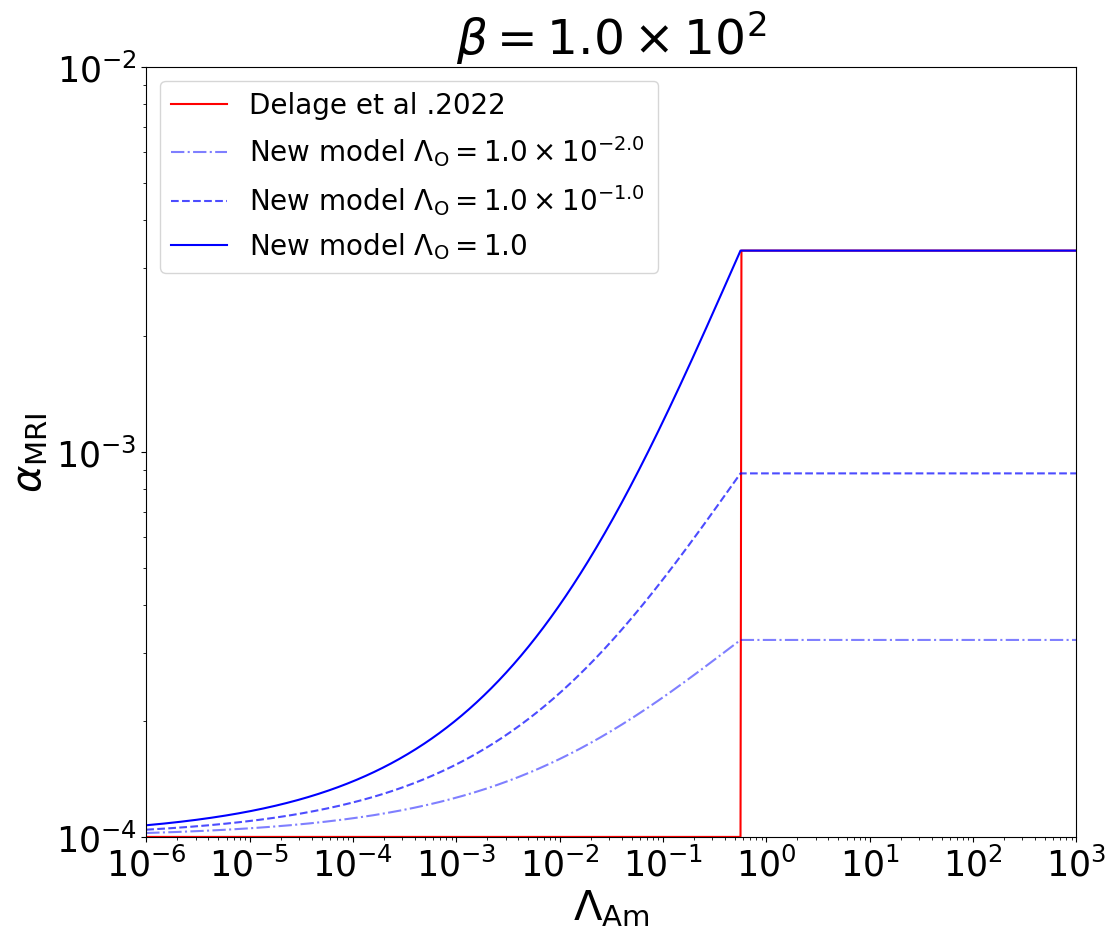}
  \caption{
Turbulent viscosity $\alpha_{\mathrm{MRI}}$ calculated using the 
smooth approximation given by Equation~(\ref{eq:alpha_mri_smooth}), 
as a function of the ambipolar Elsasser number 
$\Lambda_{\mathrm{Am}}$ for $\beta=1.0\times10^2$ and three fixed 
values of the Ohmic Elsasser number $\Lambda_O$. 
The dash-dotted, dashed, and solid blue curves correspond to 
$\Lambda_O=10^{-2}$, $10^{-1}$, and $1.0$, respectively. 
The step-function prescription adopted by \citet{Delage2022} is 
also shown as the thick red line. 
All curves approach the floor value 
$\alpha_{\mathrm{hydro}}=1.0\times10^{-4}$ at small 
$\Lambda_{\mathrm{Am}}$ and asymptotically approach plateau values 
that increase with $\Lambda_O$.
}
  \label{fig:alpha_model}
\end{figure}

When the MRI is inactive throughout the vertical extent of the disk a given radius $r^{\prime}$, we refer to that region as a dead zone. 
This condition corresponds to
\begin{align}
\bar{\alpha}(r^{\prime}) \simeq \alpha_{\mathrm{hydro}}.
\label{eq:dead_zone}
\end{align}
Figure~\ref{fig:alpha_model} shows $\alpha_{\mathrm{MRI}}$ calculated using the smooth approximation given by Equation~(\ref{eq:alpha_mri_smooth}) for $\beta = 1.0\times10^2$ as a function of $\Lambda_{\mathrm{Am}}$, considering three fixed values of $\Lambda_O = 10^{-2}$, $10^{-1}$, and $1.0$. 
In all three cases, $\alpha_{\mathrm{MRI}}$ approaches the floor value $\alpha_{\mathrm{hydro}}=1.0\times10^{-4}$ at small $\Lambda_{\mathrm{Am}}$ and increases monotonically with $\Lambda_{\mathrm{Am}}$, asymptotically approaching a value that increases with $\Lambda_O$.
The asymptotic value reaches $\alpha_{\mathrm{MRI}} \sim 3\times10^{-4}$ for $\Lambda_O = 10^{-2}$, $\sim 9\times10^{-4}$ for $\Lambda_O = 10^{-1}$, and $\sim 4\times10^{-3}$ for $\Lambda_O = 1.0$, the last of which is close to $\alpha_{\max}=1/(3\beta) \simeq 3.3\times10^{-3}$.
For $\Lambda_O = 1.0$, the step-function prescription of \citet{Delage2022} jumps from $\alpha_{\mathrm{hydro}}$ to $\alpha_{\max}$ at $\Lambda_{\mathrm{Am}} \simeq 0.5$, whereas the continous approximation introduced in this work transitions gradually over several orders of magnitude in $\Lambda_{\mathrm{Am}}$.

\subsection{Numerical settings}
\label{subsec:numerical settings}
At each timestep, we first obtain the vertically averaged magnetic 
diffusivities $\bar{\eta}_O(r,t)$ and $\bar{\eta}_A(r,t)$ 
from Equations~(\ref{eq:etaObar}) and 
(\ref{eq:etaAbar}) and substitute them into the 
induction equation (Equation~(\ref{eq:induction})) 
to update $B_z(r,t)$. 
The MRI-driven turbulent viscosity $\bar{\alpha}(r,t)$ 
is then evaluated from $B_z(r,t)$ and the magnetic 
diffusivities. Using $\bar{\alpha}$, we calculate the relative 
velocities between dust grains 
(Equations~(\ref{eq:VT})--(\ref{eq:VD})), evaluate the coagulation 
and fragmentation kernels $\bar{K}$ and $\bar{F}$ 
(Equations~(\ref{eq:K}) and (\ref{eq:F})), and update the dust size 
distribution $\Sigma_d(m)$ via Equations~(\ref{eq:dust_sigma}) and 
(\ref{eq:growthterm}).

The computational domain spans $r = 1$--$100\,\mathrm{au}$ in the 
radial direction, discretized into 200 logarithmically spaced bins. 
Assuming symmetry about the disk midplane, the vertical domain spans 
$z = 0$--$3H_g$, discretized into 300 uniformly spaced bins. 
The dust size range spans $a_d = 10^{-5}$--$10^{3}\,\mathrm{cm}$, 
discretized into 200 bins following the method described in 
Appendix A of \citet{Kawasaki2022}.

We adopt zero-gradient boundary conditions at both the inner and outer 
edges of the disk for the dust surface density. 
For the magnetic field strength, a zero-gradient boundary condition 
is adopted at the inner boundary, while the magnetic field strength 
at the outer boundary is fixed at 
$B_{z,b} = 1.0\times10^{-5}\,\mathrm{G}$ throughout the calculation.
The initial distribution of turbulent viscosity is determined 
self-consistently by iterating between the dust size distribution, 
dust surface density, and magnetic field until convergence is achieved.

We explore the parameter space by varying the initial midplane plasma 
beta $\beta_{\mathrm{ini}}$ and the fragmentation velocity 
$v_{\mathrm{frag}}$. The models considered in this study are summarized 
in Table~\ref{tab:models}. 
The fiducial model is b3-frg30, which has an initial plasma beta of 
$\beta_{\mathrm{ini}} = 10^3$ and a fragmentation velocity of 
$v_{\mathrm{frag}} = 30\ \mathrm{m\ s^{-1}}$, a value appropriate 
for icy grains. 
To investigate the dependence on the magnetic field strength, we also 
consider models b2-frg30 and b4-frg30, which have stronger 
($\beta_{\mathrm{ini}} = 10^2$) and weaker 
($\beta_{\mathrm{ini}} = 10^4$) initial magnetic fields, respectively, 
while keeping $v_{\mathrm{frag}}$ fixed. 
To investigate the dependence on the fragmentation velocity, we 
consider model b3-frg3, which has the same 
$\beta_{\mathrm{ini}} = 10^3$ as the fiducial model but adopts 
$v_{\mathrm{frag}} = 3\ \mathrm{m\ s^{-1}}$, a value representative 
of silicate grains.

\begin{table*}[htbp]
\centering
\setlength{\tabcolsep}{20pt}
\caption{
Summary of model parameters. Model b2-frg30 is the strong-magnetic-field model, b3-frg30 is the fiducial model, b4-frg30 is the weak-magnetic-field model, and b3-frg3 is the low-fragmentation-velocity model.
}
\label{tab:models}
\begin{tabular}{ccc}
\toprule
Model & $v_\mathrm{frag}\ [\mathrm{m\,s^{-1}}]$ 
& $\beta_\mathrm{ini}$ \\
\midrule
b2-frg30 & $30$ & $10^2$ \\
b3-frg30 & $30$ & $10^3$ \\
b4-frg30 & $30$ & $10^4$ \\
b3-frg3  & $3$  & $10^3$ \\
\bottomrule
\end{tabular}
\end{table*}

\section{Results} \label{sec:results}
In this section, we present the results of our calculations. 
We first describe the fiducial model (b3-frg30) in detail, focusing on the time evolution of the magnetic field, MRI-driven turbulence, dust growth, and magnetic diffusivities. 
We then present the results for models with different initial plasma beta values (b2-frg30 and b4-frg30) and for a model with a lower fragmentation velocity (b3-frg3).

\subsection{Fiducial Model: b3-frg30} 
\label{subsec:fiducial model}
\subsubsection{Time evolution of physical quantities}
\begin{figure*}[t]
  \centering
  \includegraphics[width=0.7\textwidth]
  {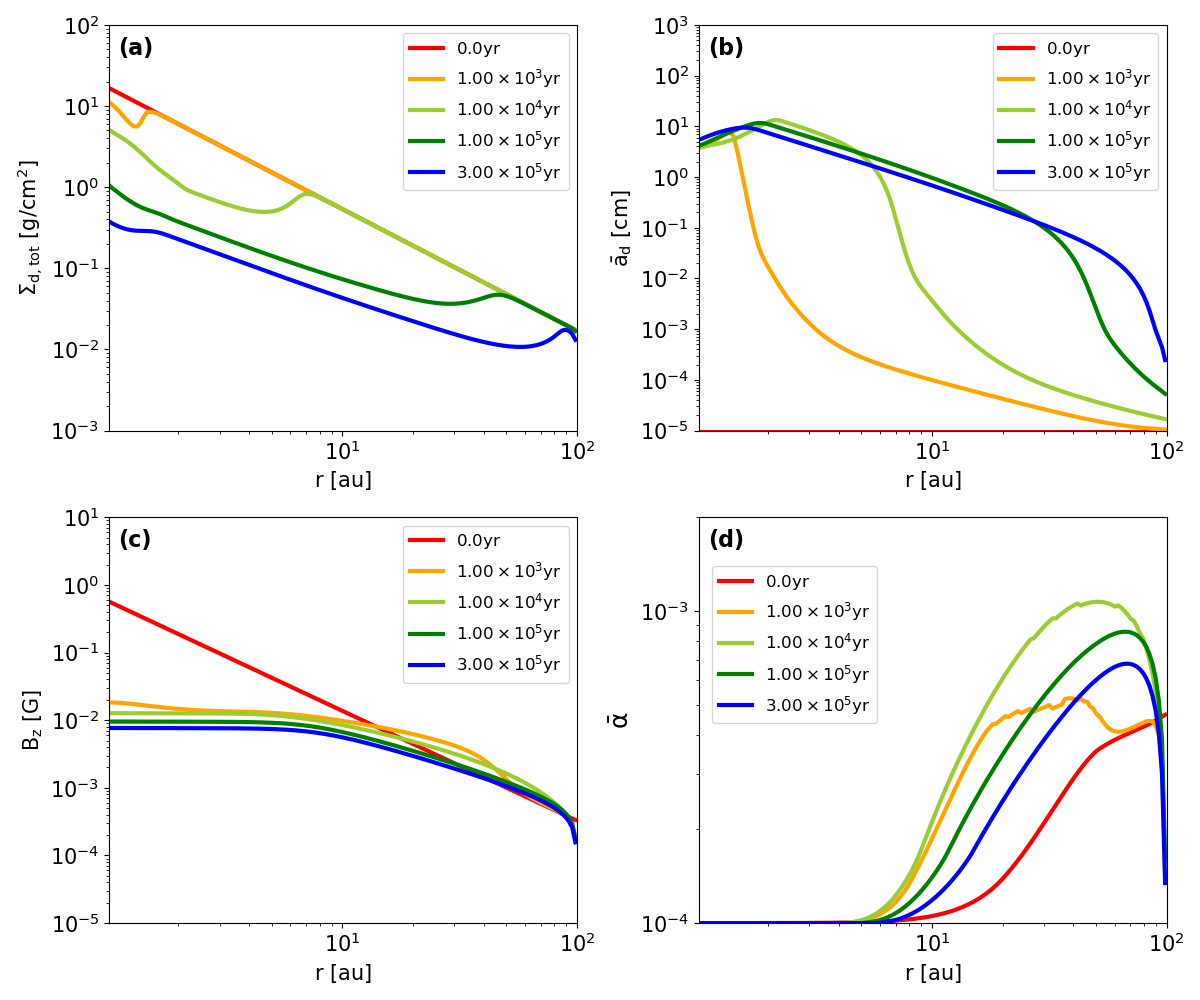}
  \caption{
  Time evolution of physical quantities for the 
fiducial model (b3-frg30). 
(a) dust surface density $\Sigma_\mathrm{d,tot}$, 
(b) mean dust size $\bar{a}_d$, 
(c) vertical magnetic field strength $B_z$, and 
(d) vertically averaged turbulent 
viscosity $\bar{\alpha}$. 
Different colors correspond to different epochs: 
$t = 0$ (red), $1.0\times10^3\,\mathrm{yr}$ 
(orange), $1.0\times10^4\,\mathrm{yr}$ (yellow-green), 
$1.0\times10^5\,\mathrm{yr}$ (green), and 
$3.0\times10^5\,\mathrm{yr}$ (blue).
  }
  \label{fig:fiducial_evol}
\end{figure*}

Figure~\ref{fig:fiducial_evol} shows the time evolution of the dust surface density, the mean dust size, the magnetic field strength, and the vertically averaged turbulent viscosity. 
The mean dust size is defined as
\begin{align}
\bar{a}_d = \frac{1}{\Sigma_{d,\mathrm{tot}}}
\int_{0}^{\infty} a_d \frac{d\Sigma_d(m)}{dm} \mathrm{d}m.
\end{align}
The dust surface density (Figure~\ref{fig:fiducial_evol}a) decreases with time. Depletion first develops in the inner region of the disk and subsequently spreads outward; by $t = 3.0\times10^5\,\mathrm{yr}$, significant depletion extends to $r \lesssim 90\,\mathrm{au}$, reflecting efficient radial drift over most of the disk.

An exception is found at $r \lesssim 1.4\,\mathrm{au}$ for $t \gtrsim 10^4\,\mathrm{yr}$, where the reduction in dust surface density is more gradual than in the surrounding regions. In this region, collision velocities frequently exceed the fragmentation velocity, preventing efficient dust growth. As a result, the grains remain relatively small and radial drift is inefficient, allowing the dust surface density to remain relatively high.

The mean dust size (Figure~\ref{fig:fiducial_evol}b) shows a radial profile that peaks at $\bar{a}_d \simeq 9.4\,\mathrm{cm}$ near $r \simeq 1.8\,\mathrm{au}$ at $t = 3.0\times10^5\,\mathrm{yr}$. The mean dust size decreases both inward and outward from this location. At $r \simeq 1.0\,\mathrm{au}$, fragmentation limits the mean dust size to $\bar{a}_d \simeq 3\,\mathrm{cm}$. At $r \simeq 1.0\times10^2\,\mathrm{au}$, by contrast, the mean dust size drops to $\bar{a}_d \simeq 2.0\times10^{-4}\,\mathrm{cm}$ owing to inefficient growth in the outer region of the disk.

The magnetic field (Figure~\ref{fig:fiducial_evol}c) diffuses outward from $r \lesssim 10\,\mathrm{au}$ to $r \gtrsim 10\,\mathrm{au}$ during the early evolution ($t < 1.0\times10^3\,\mathrm{yr}$). After this phase, the radial distribution of the magnetic field strength remains nearly unchanged, indicating that magnetic diffusion becomes negligible thereafter.

The turbulent viscosity (Figure~\ref{fig:fiducial_evol}d) initially increases monotonically with radius at $r \gtrsim 10\,\mathrm{au}$, reaching a maximum value of $\bar{\alpha} \simeq 4.2\times10^{-4}$ near the outer boundary ($r \simeq 100\,\mathrm{au}$) at $t=0$. 
At $t = 1.0\times10^3\,\mathrm{yr}$, the peak shifts inward to $r \simeq 50\,\mathrm{au}$, with a value of $\bar{\alpha} \simeq 1.0\times10^{-3}$. For $t \geq 1.0\times10^4\,\mathrm{yr}$, the peak location moves outward while its magnitude gradually decreases.

\subsubsection{Dust size distribution}
\begin{figure*}[t]
  \centering
  \includegraphics[width=0.7\textwidth]
  {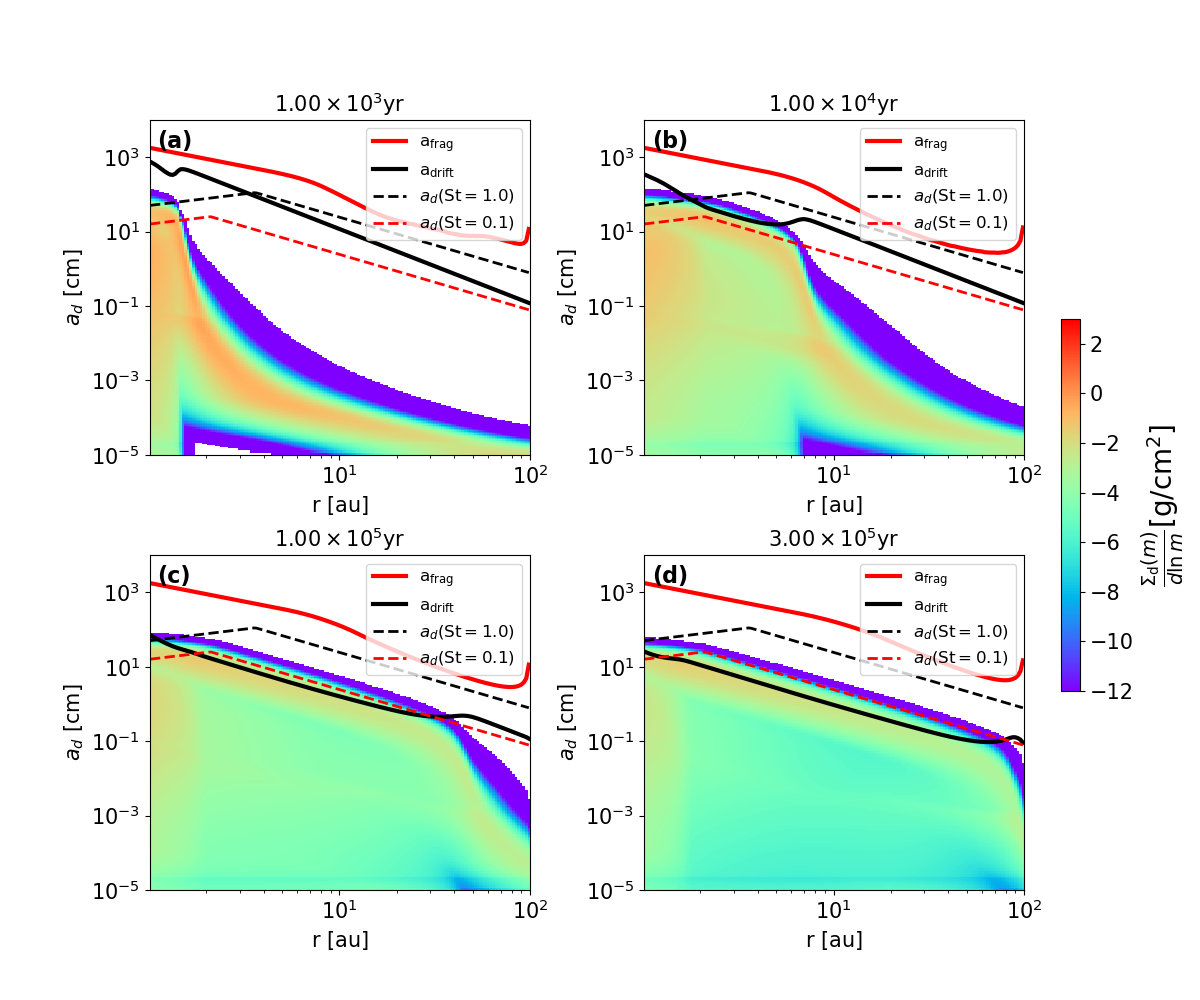}
  \caption{
  Distribution of the dust surface density
$\Sigma_d(m)$ as a function of dust size
$a_d$ and orbital radius $r$ at different epochs
for the fiducial model (b3-frg30).
The black and red solid lines show the
drift-limited size $a_{\mathrm{drift}}$ and
fragmentation-limited size $a_{\mathrm{frag}}$,
respectively \citep{Birnstiel2012}.
The black and red dashed lines indicate the dust
sizes $a_d(\mathrm{St}=1)$ and
$a_d(\mathrm{St}=0.1)$,
respectively.
The panels correspond to
(a) $t = 1.0\times10^3\,\mathrm{yr}$,
(b) $1.0\times10^4\,\mathrm{yr}$,
(c) $1.0\times10^5\,\mathrm{yr}$,
and (d) $3.0\times10^5\,\mathrm{yr}$.
  }
  \label{fig:dsize}
\end{figure*}

Figure~\ref{fig:dsize} shows the distribution of the dust surface density as a function of dust size $a_d$ and radius $r$ at different epochs. The drift-limited size $a_{\mathrm{drift}}$ and fragmentation-limited size $a_{\mathrm{frag}}$ are shown by the black and red solid lines, respectively, following \citet{Birnstiel2012}:
\begin{align}
a_{\mathrm{frag}} &=
f_f\,\frac{2\Sigma_g v_{\mathrm{frag}}^2}
{3\pi\rho_{\mathrm{in}}\bar{\alpha} c_s^2}, \\
a_{\mathrm{drift}} &=
f_d\,\frac{2\Sigma_g r^2\Omega^2}
{\pi\rho_{\mathrm{in}} c_s^2}
\left|\frac{\mathrm{d}\ln P}{\mathrm{d}\ln r}
\right|^{-1},
\end{align}
where $f_f = 0.77$ and $f_d = 0.55$ are numerical factors adopted from \citet{Birnstiel2012}. We note that $a_{\mathrm{drift}} < a_{\mathrm{frag}}$ holds throughout most of the disk at all epochs in this model.

At $t = 1.0\times10^3\,\mathrm{yr}$ (Figure~\ref{fig:dsize}a), dust grows to $\simeq 50\,\mathrm{cm}$ at $r \lesssim 1.5\,\mathrm{au}$. Grains of this size have collision velocities high enough to exceed $v_{\mathrm{frag}}$, leading to efficient fragmentation and replenishment of small grains around the monomer size ($a_d \sim a_0 = 1.0\times10^{-5}\,\mathrm{cm}$) in this region. By contrast, in the region $1.5\,\mathrm{au} \lesssim r \lesssim 10\,\mathrm{au}$, small grains are depleted as coagulation dominates over fragmentation.

At $t = 1.0\times10^4\,\mathrm{yr}$ (Figure~\ref{fig:dsize}b), in the range $2.0\,\mathrm{au} \lesssim r \lesssim 6.0\,\mathrm{au}$, the maximum dust size (defined as the largest $a_d$ satisfying $\Sigma_d(a_d) > 1.0\times10^{-12}\,\mathrm{g\,cm^{-2}}$) has reached $a_{\mathrm{drift}}$. Even in this drift-dominated region, fragmentation continually replenishes grains around the monomer size ($a_d \sim a_0$). Consequently, small grains remain abundant despite the continuous loss of larger grains by radial drift after they reach $a_{\mathrm{drift}}$.

At $t = 1.0\times10^5\,\mathrm{yr}$ (Figure~\ref{fig:dsize}c), the maximum dust size slightly exceeds $a_{\mathrm{drift}}$ in the region $r \lesssim 40\,\mathrm{au}$, indicating that dust growth has reached the drift barrier in this region. At $r \simeq 40\,\mathrm{au}$, the maximum dust size is $\simeq 1.0\,\mathrm{cm}$. By $t = 3.0\times10^5\,\mathrm{yr}$ (Figure~\ref{fig:dsize}d), the region in which dust growth has reached the drift barrier extends to $r \simeq 80\,\mathrm{au}$. At $r \simeq 1.0\times10^2\,\mathrm{au}$, by contrast, the maximum dust size remains as small as $\simeq 0.1\,\mathrm{cm}$. In the region $2.2\,\mathrm{au} \lesssim r \lesssim 80\,\mathrm{au}$, the dust size corresponding to $\mathrm{St}=0.1$ (red dashed line) approximately coincides with $a_{\mathrm{drift}}$, indicating that radial drift becomes efficient as grains approach $\mathrm{St}\sim0.1$.

\begin{figure*}[t]
  \centering
  \includegraphics[width=\linewidth]
  {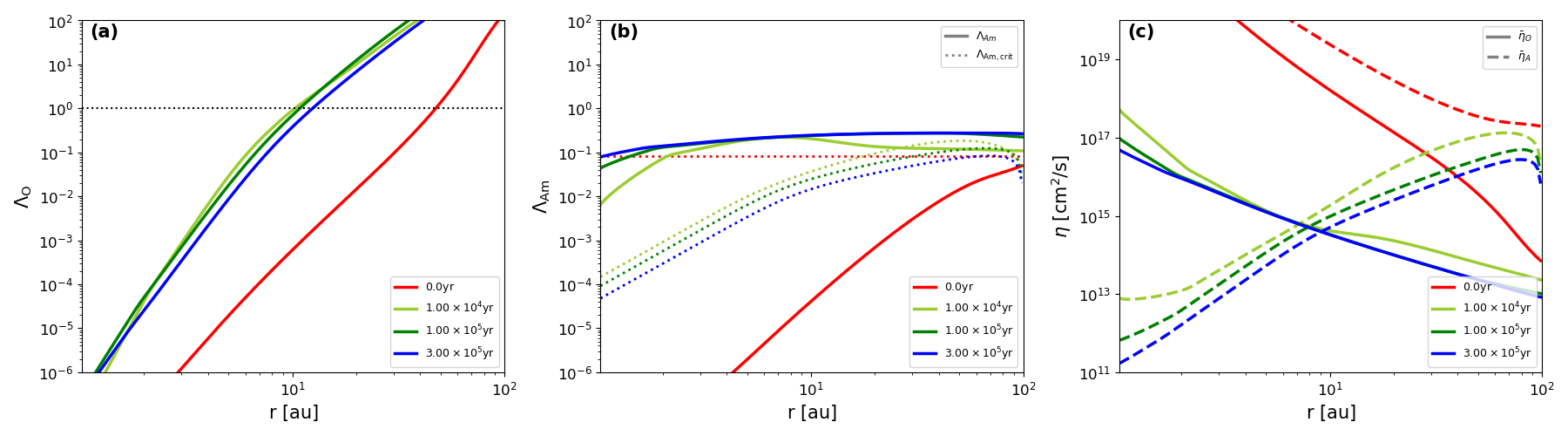}
  \caption{
  Radial profiles of the Elsasser numbers and magnetic diffusivities 
at different epochs for the fiducial model (b3-frg30). 
Panel (a) shows the Ohmic Elsasser number $\Lambda_O$ (solid lines) 
and the threshold $\Lambda_O = 1$ (dotted line), above which 
Ohmic resistivity does not suppress the MRI. 
Panel (b) shows the ambipolar Elsasser number $\Lambda_{\mathrm{Am}}$ 
(solid lines) and its critical value $\Lambda_{\mathrm{Am,crit}}$ 
(dotted lines), above which ambipolar diffusion does not suppress 
the MRI. 
Panel (c) shows the vertically averaged magnetic diffusivities 
for Ohmic resistivity $\bar{\eta}_O$ (solid lines) and 
ambipolar diffusion $\bar{\eta}_A$ (dashed lines). 
Different colors correspond to different epochs: 
$0\,\mathrm{yr}$ (red), 
$1.0\times10^4\,\mathrm{yr}$ (yellow-green), 
$1.0\times10^5\,\mathrm{yr}$ (green), and 
$3.0\times10^5\,\mathrm{yr}$ (blue).
  }
  \label{fig:eta}
\end{figure*}

\subsubsection{Magnetic diffusivities}

$\Lambda_{\rm Am,crit}$ is the quantity introduced in
Section~\ref{subsec:model of MRI} (Equation~(\ref{eq:lambda_crit})).
The condition $\Lambda_{\mathrm{Am}} > \Lambda_{\mathrm{Am,crit}}$ is
therefore equivalent to $\beta < \beta_{\min}(\Lambda_{\mathrm{Am}})$
(Equation~(\ref{eq:beta_condition})).

Figure~\ref{fig:eta} shows the radial profiles of 
(a) the Ohmic Elsasser number $\Lambda_O$, 
(b) the ambipolar Elsasser number $\Lambda_{\mathrm{Am}}$ together 
with $\Lambda_{\mathrm{Am,crit}}$, and 
(c) the vertically averaged magnetic diffusivities $\bar{\eta}_O$ 
and $\bar{\eta}_A$. 
In panel (a), $\Lambda_O$ increases monotonically with radius at all 
epochs and, for $t \gtrsim 1.0\times10^4\,\mathrm{yr}$, crosses unity near $r \simeq 10\,\mathrm{au}$. 
In panel (b), $\Lambda_{\mathrm{Am}}$ also increases with radius before leveling off in the outer disk. 
At $t \gtrsim 1.0\times10^4\,\mathrm{yr}$, $\Lambda_{\mathrm{Am}}$ already exceeds $\Lambda_{\mathrm{Am,crit}}$ within $r \lesssim 10\,\mathrm{au}$.
Panel (c) shows that, at $t \gtrsim 1.0\times10^4\,\mathrm{yr}$, 
Ohmic resistivity dominates in the inner disk 
($r \lesssim 10\,\mathrm{au}$; $\bar{\eta}_O > \bar{\eta}_A$), 
whereas ambipolar diffusion dominates in the outer disk 
($r \gtrsim 10\,\mathrm{au}$; $\bar{\eta}_A > \bar{\eta}_O$).

In both regions, the decrease in magnetic diffusivity over time is 
primarily caused by dust growth, which depletes small grains and thereby 
increases the abundance of charged particles. 
Consequently, $\bar{\eta}_O$ decreases in the inner disk, where Ohmic 
resistivity dominates, while $\bar{\eta}_A$ decreases in the outer disk, 
where ambipolar diffusion dominates. 
The change in $B_z$ is relatively small at $r \gtrsim 10\,\mathrm{au}$ 
and therefore makes only a minor contribution to the evolution of 
$\bar{\eta}_A$.

The regions where MRI-driven turbulence is active
($\bar{\alpha} > \alpha_{\mathrm{hydro}}$ in
Figure~\ref{fig:fiducial_evol}d) broadly correspond, at each epoch, 
to the regions satisfying both $\Lambda_O \gtrsim 1$ 
(Figure~\ref{fig:eta}a) and 
$\Lambda_{\mathrm{Am}} \gtrsim \Lambda_{\mathrm{Am,crit}}$
(Figure~\ref{fig:eta}b). 
By $t \gtrsim 1.0\times10^4\,\mathrm{yr}$, both conditions are satisfied over most of $r \gtrsim 10\,\mathrm{au}$, and the MRI becomes active there. 
At $t = 1.0\times10^4\,\mathrm{yr}$, $\Lambda_{\mathrm{Am}}$
falls below $\Lambda_{\mathrm{Am,crit}}$ within
$30\,\mathrm{au} \lesssim r \lesssim 80\,\mathrm{au}$. 
Because $\alpha_{\mathrm{MRI}}$ depends smoothly on 
$\Lambda_{\mathrm{Am}}$ (Equation~(\ref{eq:alpha_mri_smooth})), 
the turbulent viscosity remains above $\alpha_{\mathrm{hydro}}$ 
in this range, consistent with the sustained MRI activity seen in Figure~\ref{fig:fiducial_evol}d. 
The evolution of the MRI-active region is therefore controlled by the weakening of both Ohmic resistivity and ambipolar diffusion as dust grains grow.

\label{subsec:strong_model}
\begin{figure*}[t]
  \centering
  \includegraphics[width=0.7\textwidth]
  {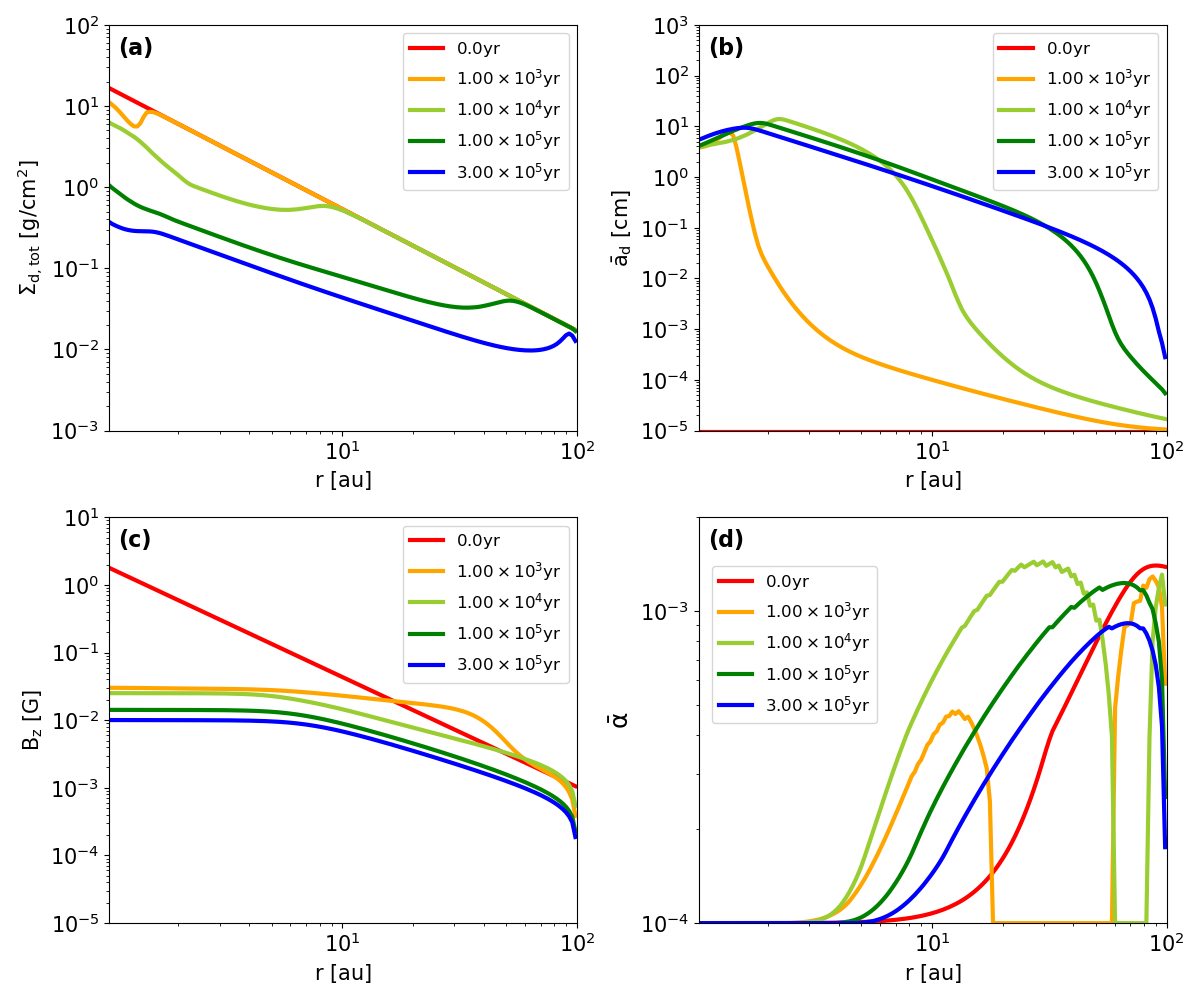}
  \caption{
    Same as Figure~\ref{fig:fiducial_evol}, but 
    for the strong-magnetic-field model 
    (b2-frg30) with $\beta_{\mathrm{ini}} = 10^2$.
  }
  \label{fig:beta2_evol}
\end{figure*}
\begin{figure*}[t]
  \centering
  \includegraphics[width=\linewidth]
  {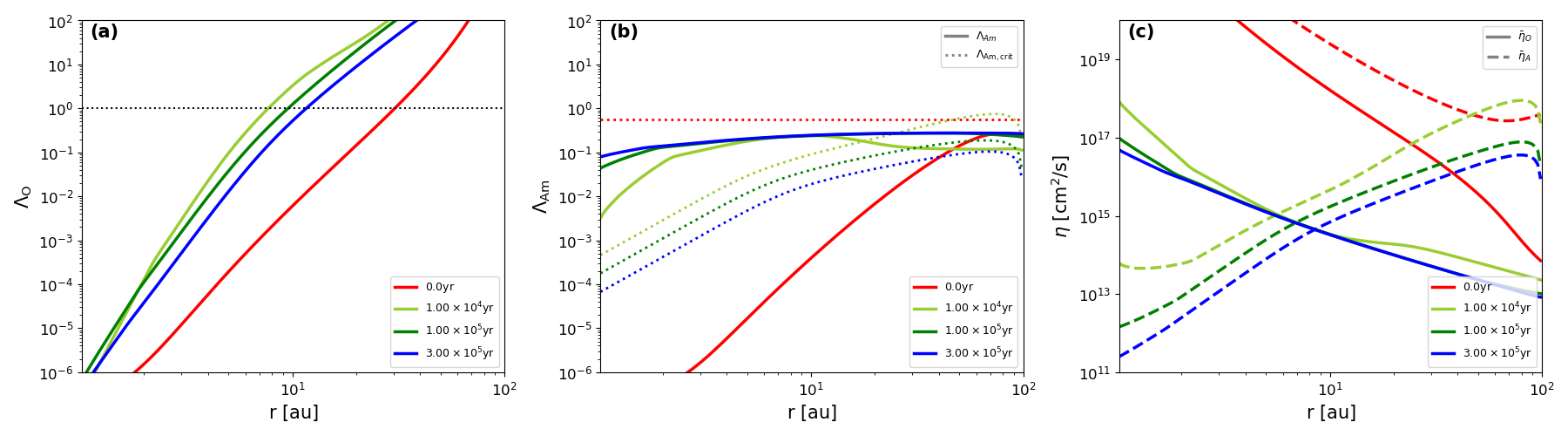}
  \caption{
    Same as Figure~\ref{fig:eta}, but 
    for the strong-magnetic-field model 
    (b2-frg30) with $\beta_{\mathrm{ini}} = 10^2$.
  }
  \label{fig:eta_strong}
\end{figure*}

\subsection{Strong-Magnetic-Field Model: b2-frg30}
We next examine the strong-magnetic-field model (b2-frg30), which has a stronger initial magnetic field ($\beta_{\mathrm{ini}} =
10^2$) than the fiducial model (b3-frg30; $\beta_{\mathrm{ini}} =
10^3$). Figure~\ref{fig:beta2_evol} shows the time evolution of the same physical quantities as in Figure~\ref{fig:fiducial_evol}.
The dust surface density (Figure~\ref{fig:beta2_evol}a), mean dust size (Figure~\ref{fig:beta2_evol}b), and magnetic field strength (Figure~\ref{fig:beta2_evol}c) evolve similarly to those in the fiducial model. The turbulent viscosity (Figure~\ref{fig:beta2_evol}d), however, shows a distinct evolution.

Figure~\ref{fig:eta_strong} shows the Elsasser numbers and magnetic diffusivities for this model, as in Figure~\ref{fig:eta}. 
The Ohmic Elsasser number (Figure~\ref{fig:eta_strong}a) and the magnetic diffusivities (Figure~\ref{fig:eta_strong}c) evolve similarly to those in the fiducial model, whereas the ambipolar Elsasser number (Figure~\ref{fig:eta_strong}b) shows a distinct evolution.

At $t = 0$, the turbulent viscosity (Figure~\ref{fig:beta2_evol}d)
reaches a peak value of $\bar{\alpha} \simeq 1.0\times10^{-3}$ at
$r \simeq 1.0\times10^2\,\mathrm{au}$, more than twice the peak value in the fiducial model ($\bar{\alpha} \simeq 4.2\times10^{-4}$; Section~\ref{subsec:fiducial model}). This enhancement follows directly from the stronger initial magnetic field through the relation $\alpha_{\mathrm{MRI}} \sim 1/\beta$ (Equation~(\ref{eq:alpha_mri_smooth})).

However, at $t = 1.0\times10^4\,\mathrm{yr}$, MRI-driven turbulence is suppressed in the region $60\,\mathrm{au} \lesssim r \lesssim 80\,\mathrm{au}$ (Figure~\ref{fig:beta2_evol}d). As the magnetic field diffuses outward from the inner disk, $B_z$ in this region increases significantly. Because $\bar{\eta}_A$ increases with magnetic field strength (Equation~(\ref{eq:eta_A})), this increase in $B_z$ strengthens ambipolar diffusion enough to suppress the MRI.  Figure~\ref{fig:eta_strong}b shows that $\Lambda_{\mathrm{Am}}$ falls below $\Lambda_{\mathrm{Am,crit}}$ in this region at $t = 1.0\times10^4\,\mathrm{yr}$, whereas $\Lambda_O$ remains well above unity throughout the same region (Figure~\ref{fig:eta_strong}a). The dead zone is therefore caused by ambipolar diffusion rather than Ohmic resistivity.

At later epochs ($t \gtrsim 1.0\times10^5\,\mathrm{yr}$), the MRI becomes active again in $60\,\mathrm{au} \lesssim r \lesssim 80\,\mathrm{au}$, with $\bar{\alpha} \simeq 1.0\times10^{-3}$ (Figure~\ref{fig:beta2_evol}d). As the magnetic field continues to diffuse outward, the local magnetic field strength decreases, weakening ambipolar diffusion. Consistently, Figure~\ref{fig:eta_strong}b shows that $\Lambda_{\mathrm{Am}}$ rises above $\Lambda_{\mathrm{Am,crit}}$ in this region at these epochs.

Compared with the fiducial model (b3-frg30), this model is characterized by stronger turbulence in the outer disk. It also exhibits the transient formation of a dead zone in the region $60\,\mathrm{au} \lesssim r \lesssim 80\,\mathrm{au}$, caused by enhanced ambipolar diffusion associated with the transient accumulation of magnetic flux.

\label{subsec:weak_model}
\begin{figure*}[t]
  \centering
  \includegraphics[width=0.7\textwidth]
  {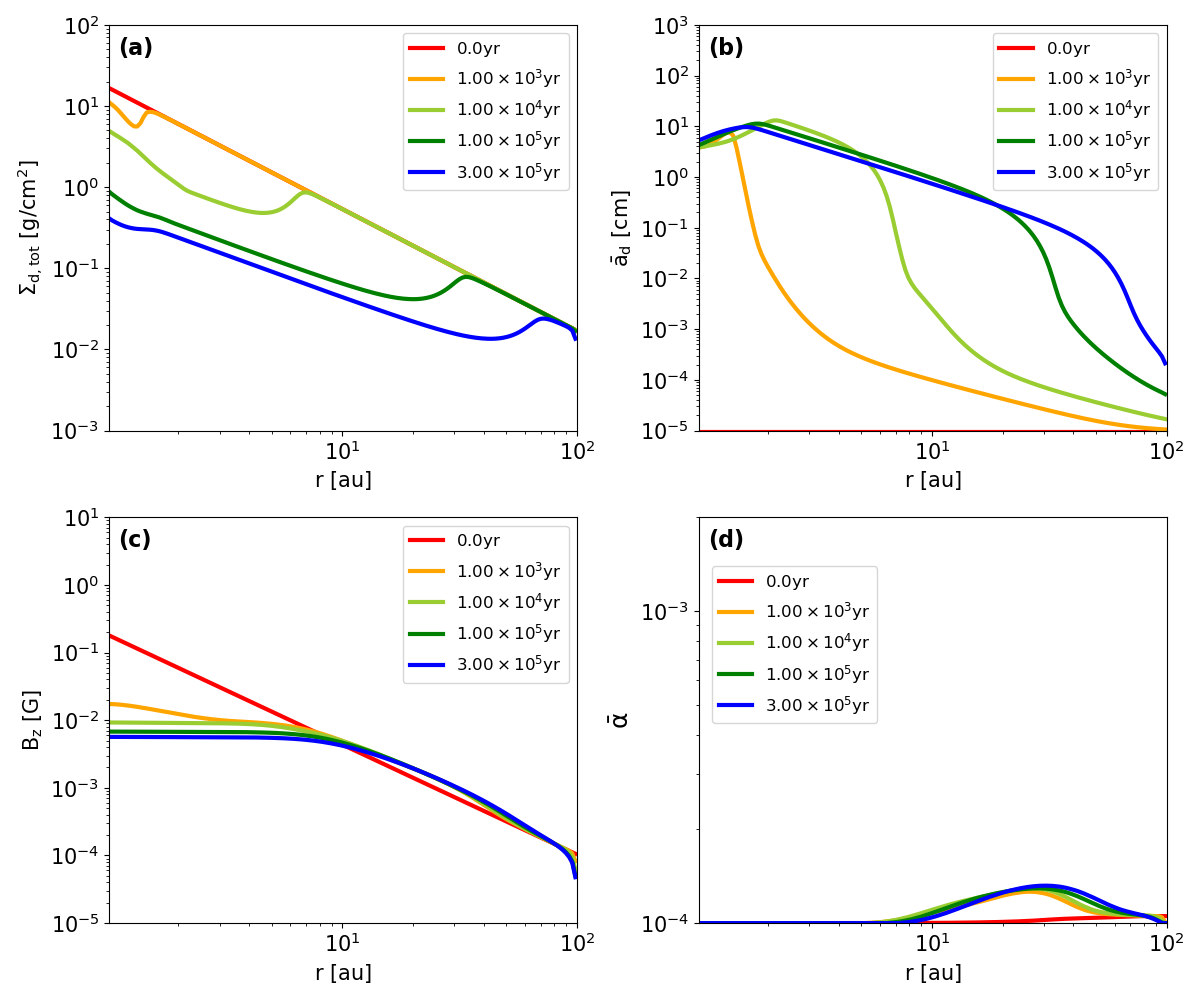}
  \caption{
    Same as Figure~\ref{fig:fiducial_evol}, but 
    for the weak-magnetic-field model 
    (b4-frg30) with $\beta_{\mathrm{ini}} = 10^4$.
  }
  \label{fig:beta4_evol}
\end{figure*}
\begin{figure*}[t]
  \centering
  \includegraphics[width=\linewidth]
  {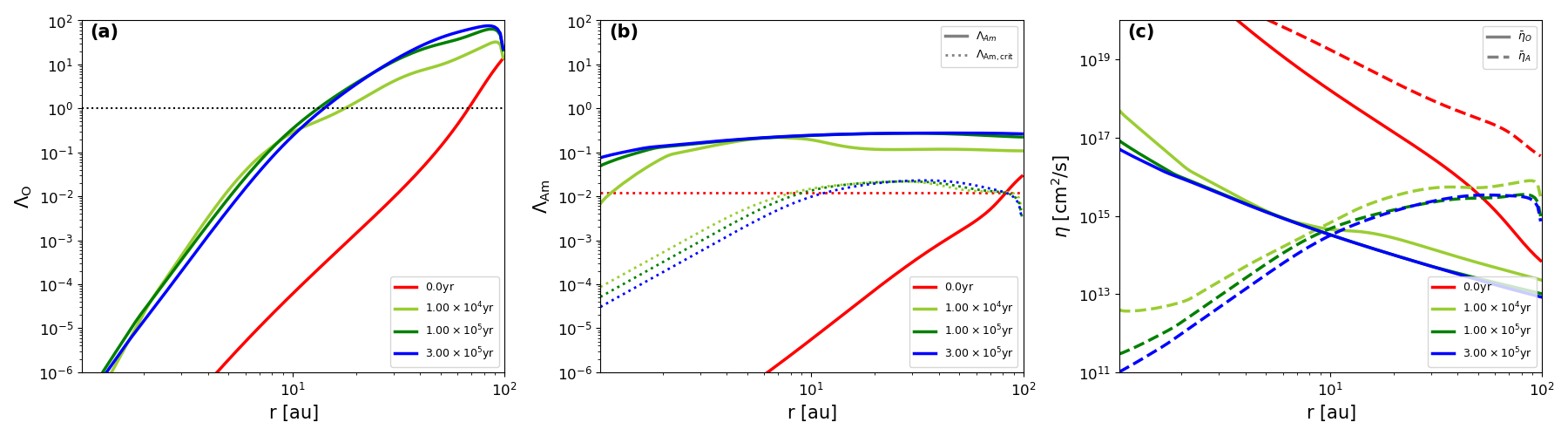}
  \caption{
    Same as Figure~\ref{fig:eta}, but 
    for the weak-magnetic-field model 
    (b4-frg30) with $\beta_{\mathrm{ini}} = 10^4$.
  }
  \label{fig:eta_weak}
\end{figure*}

\subsection{Weak-Magnetic-Field Model: b4-frg30}
We next examine the weak-magnetic-field model (b4-frg30), which has a 
weaker initial magnetic field ($\beta_{\mathrm{ini}} = 10^4$) than the 
fiducial model (b3-frg30; $\beta_{\mathrm{ini}} = 10^3$).
Figure~\ref{fig:beta4_evol} shows the time evolution of the same
physical quantities as in Figure~\ref{fig:fiducial_evol}.
The overall evolution of the dust surface density
(Figure~\ref{fig:beta4_evol}a), mean dust size
(Figure~\ref{fig:beta4_evol}b), and magnetic field strength
(Figure~\ref{fig:beta4_evol}c) is qualitatively similar to that in the fiducial model (b3-frg30).

Figure~\ref{fig:eta_weak} shows the Elsasser numbers and magnetic 
diffusivities for this model, as in Figure~\ref{fig:eta}. 
The Ohmic Elsasser number (Figure~\ref{fig:eta_weak}a) and the magnetic 
diffusivities (Figure~\ref{fig:eta_weak}c) evolve similarly to those in the fiducial model. 

The ambipolar Elsasser number (Figure~\ref{fig:beta4_evol}b), however, differs from that in the fiducial model. For $t \gtrsim 1.0 \times 10^4\,\mathrm{yr}$, $\Lambda_{\mathrm{Am}}$ remains above $\Lambda_{\mathrm{Am,crit}}$ throughout $r \gtrsim 10\,\mathrm{au}$ in this model (b4-frg30), whereas in the
fiducial model (b3-frg30) $\Lambda_{\mathrm{Am}}$ falls below
$\Lambda_{\mathrm{Am,crit}}$ in the region $30\,\mathrm{au} \lesssim r \lesssim 80\,\mathrm{au}$ at the same epochs (Figure~\ref{fig:fiducial_evol}).

The turbulent viscosity (Figure~\ref{fig:beta4_evol}d), also, shows a significant difference. 
By $t \gtrsim 1.0\times10^4\,\mathrm{yr}$, the MRI activation condition 
is satisfied over most of the disk ($\Lambda_O \gtrsim 1$ and
$\Lambda_{\mathrm{Am}} \gtrsim \Lambda_{\mathrm{Am,crit}}$ beyond
$r \simeq 10\,\mathrm{au}$), indicating that non-ideal MHD effects do 
not suppress MRI-driven turbulence there. 
Instead, because $\alpha_{\mathrm{MRI}} \sim 1/\beta$ 
(Equation~(\ref{eq:alpha_mri_smooth})), the weaker initial magnetic field in model b4-frg30 ($\beta_{\mathrm{ini}}=10^4$) yields a systematically smaller $\alpha_{\mathrm{MRI}}$ even where the activation condition is satisfied.

Because MRI-driven turbulence remains weak throughout the disk in model b4-frg30, turbulence-induced relative velocities between dust grains are lower. Dust growth via coagulation therefore proceeds more slowly than in the fiducial model (b3-frg30), and turbulence-induced fragmentation is also weak. 
In the inner disk, where dust growth reaches the drift barrier, the dust size is primarily regulated by radial drift and thus 
differs little from that in the fiducial model. 
In the outer disk, by 
contrast, dust growth has not yet reached the drift barrier, and the weaker turbulence results in slightly smaller dust sizes than in the fiducial model.

\label{subsec:low_frag_model}
\begin{figure*}[t]
  \centering
  \includegraphics[width=0.7\textwidth]
  {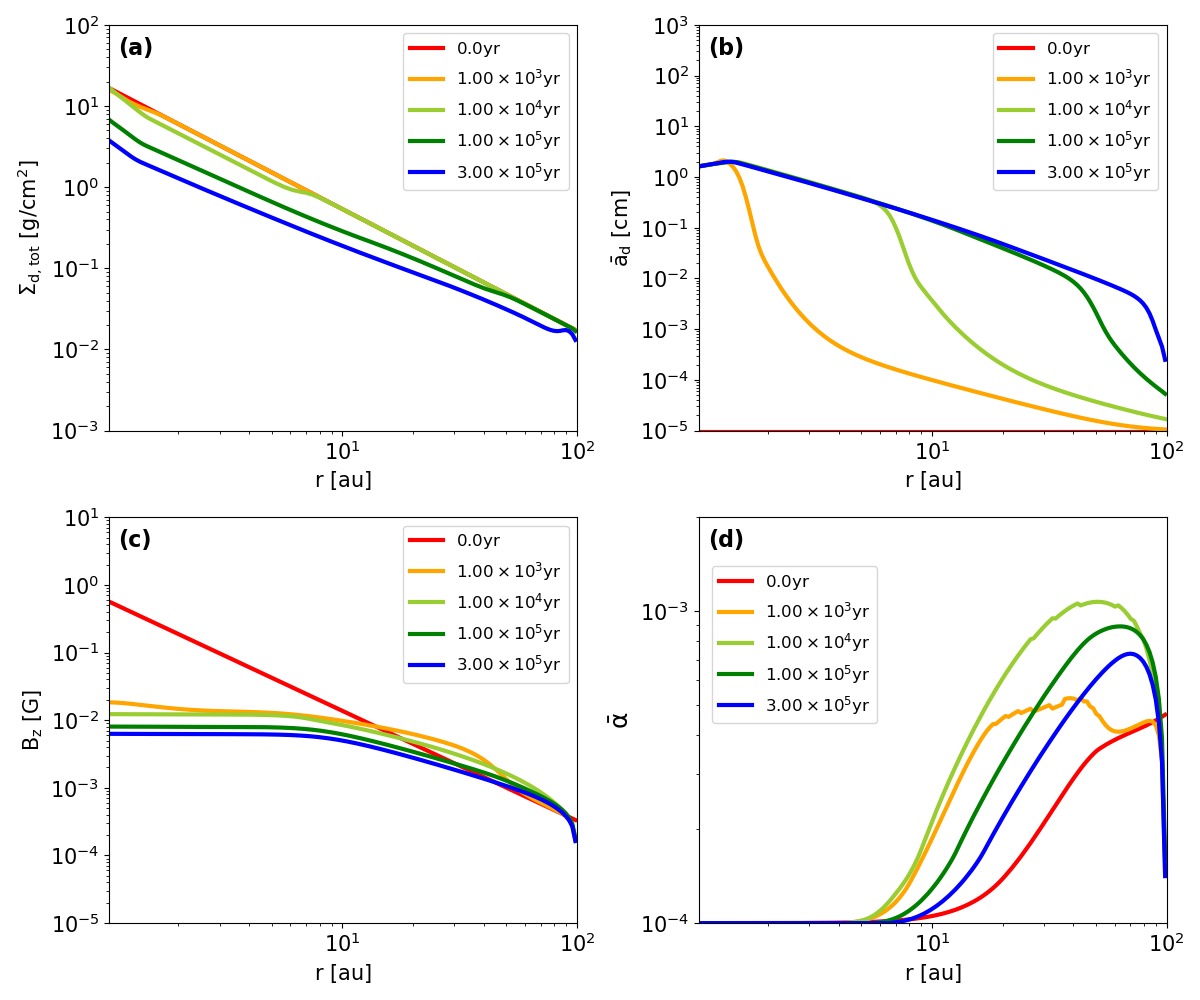}
  \caption{
    Same as Figure~\ref{fig:fiducial_evol}, but 
    for the low-fragmentation-velocity model 
    (b3-frg3) with $v_{\mathrm{frag}} = 3\,\mathrm{m\,s^{-1}}$.
  }
  \label{fig:frg3_evol}
\end{figure*}
\begin{figure*}[t]
  \centering
  \includegraphics[width=\linewidth]
  {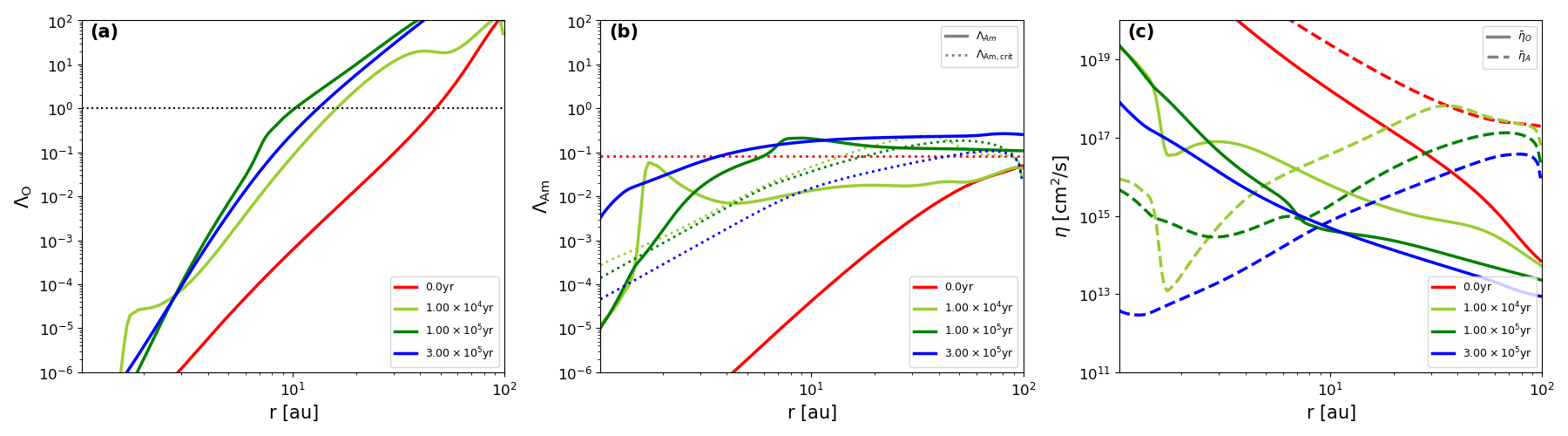}
  \caption{
    Same as Figure~\ref{fig:fiducial_evol}, but 
    for the low-fragmentation-velocity model 
    (b3-frg3) with $v_{\mathrm{frag}} = 3\,\mathrm{m\,s^{-1}}$.
  }
  \label{fig:eta_frg3}
\end{figure*}

\subsection{Low-Fragmentation-Velocity Model: b3-frg3}
We next examine the effect of a lower fragmentation velocity by 
considering model b3-frg3, which adopts 
$v_{\mathrm{frag}} = 3\,\mathrm{m\,s^{-1}}$, compared with 
$v_{\mathrm{frag}} = 30\,\mathrm{m\,s^{-1}}$ in the fiducial model 
(b3-frg30; Section~\ref{subsec:fiducial model}). 
The two models have the same initial plasma beta 
($\beta_{\mathrm{ini}} = 10^3$). 
Figure~\ref{fig:frg3_evol} shows the time evolution of the same 
physical quantities as in Figure~\ref{fig:fiducial_evol}.

The dust surface density (Figure~\ref{fig:frg3_evol}a) decreases with 
time, particularly in the inner disk, owing to efficient radial drift. 
The mean dust size (Figure~\ref{fig:frg3_evol}b), however, differs 
clearly from that in the fiducial model (b3-frg30). 
The mean dust size is significantly smaller, peaking at 
$\bar{a}_d \simeq 2\,\mathrm{cm}$ near $r \simeq 2\,\mathrm{au}$ at 
$t = 3.0\times10^5\,\mathrm{yr}$, compared with 
$\bar{a}_d \simeq 9.4\,\mathrm{cm}$ at a comparable radius 
($r \simeq 1.8\,\mathrm{au}$) in the fiducial model at the same epoch. 
This reduction arises from the lower fragmentation velocity, which 
promotes more efficient fragmentation and thereby limits grain growth 
to smaller sizes.

Figure~\ref{fig:eta_frg3} shows the Elsasser numbers and magnetic 
diffusivities for this model, as in Figure~\ref{fig:eta}. 
Figure~\ref{fig:eta_frg3}b shows a local enhancement in 
$\Lambda_{\mathrm{Am}}$ near $r \simeq 2\,\mathrm{au}$ at 
$t = 1.0\times10^4\,\mathrm{yr}$. 
Despite this local enhancement, the Ohmic Elsasser number 
(Figure~\ref{fig:eta_frg3}a) and the magnetic diffusivities 
(Figure~\ref{fig:eta_frg3}c) evolve similarly to those in the 
fiducial model. 
The magnetic field strength and turbulent viscosity 
(Figure~\ref{fig:frg3_evol}c,d) also show evolution similar to that 
in the fiducial model. 
At $t = 1.0\times10^4\,\mathrm{yr}$, the turbulent viscosity reaches 
a peak value of $\bar{\alpha} \simeq 1.0\times10^{-3}$ at 
$r \simeq 50$--$60\,\mathrm{au}$ 
(Figure~\ref{fig:frg3_evol}d), comparable to the peak value 
$\bar{\alpha} \simeq 1.0\times10^{-3}$ in the fiducial model at the 
same epoch.

\subsection{Comparison of Models}
\label{subsec:comparison}
We summarize the results of all four models to clarify the dependence 
of dust and magnetic field evolution on the initial magnetic field 
strength and fragmentation velocity.

In the fiducial model (b3-frg30), magnetic diffusion is significant at 
early epochs but becomes less important at later epochs 
(Figure~\ref{fig:eta}c), while MRI-driven turbulence gradually shifts 
outward and weakens (Figure~\ref{fig:fiducial_evol}d). 
In the strong-magnetic-field model (b2-frg30), turbulence remains 
strong in the outer disk, while a transient suppression of MRI-driven 
turbulence occurs at $60\,\mathrm{au} < r < 80\,\mathrm{au}$ owing to 
enhanced ambipolar diffusion associated with the transient accumulation 
of magnetic flux (Figure~\ref{fig:beta2_evol}). 
In contrast, in the weak-magnetic-field model (b4-frg30), MRI-driven 
turbulence remains weak even where the MRI activation conditions are 
satisfied, owing to the weaker magnetic field 
(Figure~\ref{fig:beta4_evol}). 
In the low-fragmentation-velocity model (b3-frg3), the mean dust size 
is significantly smaller than in the fiducial model owing to more 
efficient fragmentation, while the magnetic field and turbulent 
viscosity evolve in a qualitatively similar manner 
(Figure~\ref{fig:frg3_evol}).

In all models, the dust size in the inner disk is largely regulated by 
the drift barrier and therefore shows relatively little dependence on 
the initial magnetic field strength. 
In the outer disk, by contrast, where dust growth has not yet reached 
the drift barrier, the dust size depends on $\beta_{\mathrm{ini}}$, 
reflecting differences in the strength of MRI-driven turbulence among 
the models.

\section{Discussion} \label{sec:discussion}
In this section, we discuss the dependence of our results on the model parameters and the physical processes underlying the disk evolution. We first examine how the initial magnetic field strength affects the dust size distribution and magnetic diffusivities (Section~\ref{subsec:dependence_of_magnetic_field}). 
We then discuss the dependence on the fragmentation velocity 
(Section~\ref{subsec:dependence_vfrag}). 
We also compare the characteristic timescales of dust growth, radial drift, and magnetic diffusion to clarify which physical process dominates the disk evolution at each radius 
(Section~\ref{subsec:time_scale}). 

Finally, we evaluate the effects of ambipolar drift and the electrostatic barrier on dust growth in Sections~\ref{subsec:effect of ambipolar_drift} and \ref{subsec:effect of electrostatic barrier}, respectively.

\subsection{Dependence on Initial Magnetic Field Strength}
\label{subsec:dependence_of_magnetic_field}
\begin{figure*}[t]
  \centering
  \includegraphics[width=0.7\textwidth]{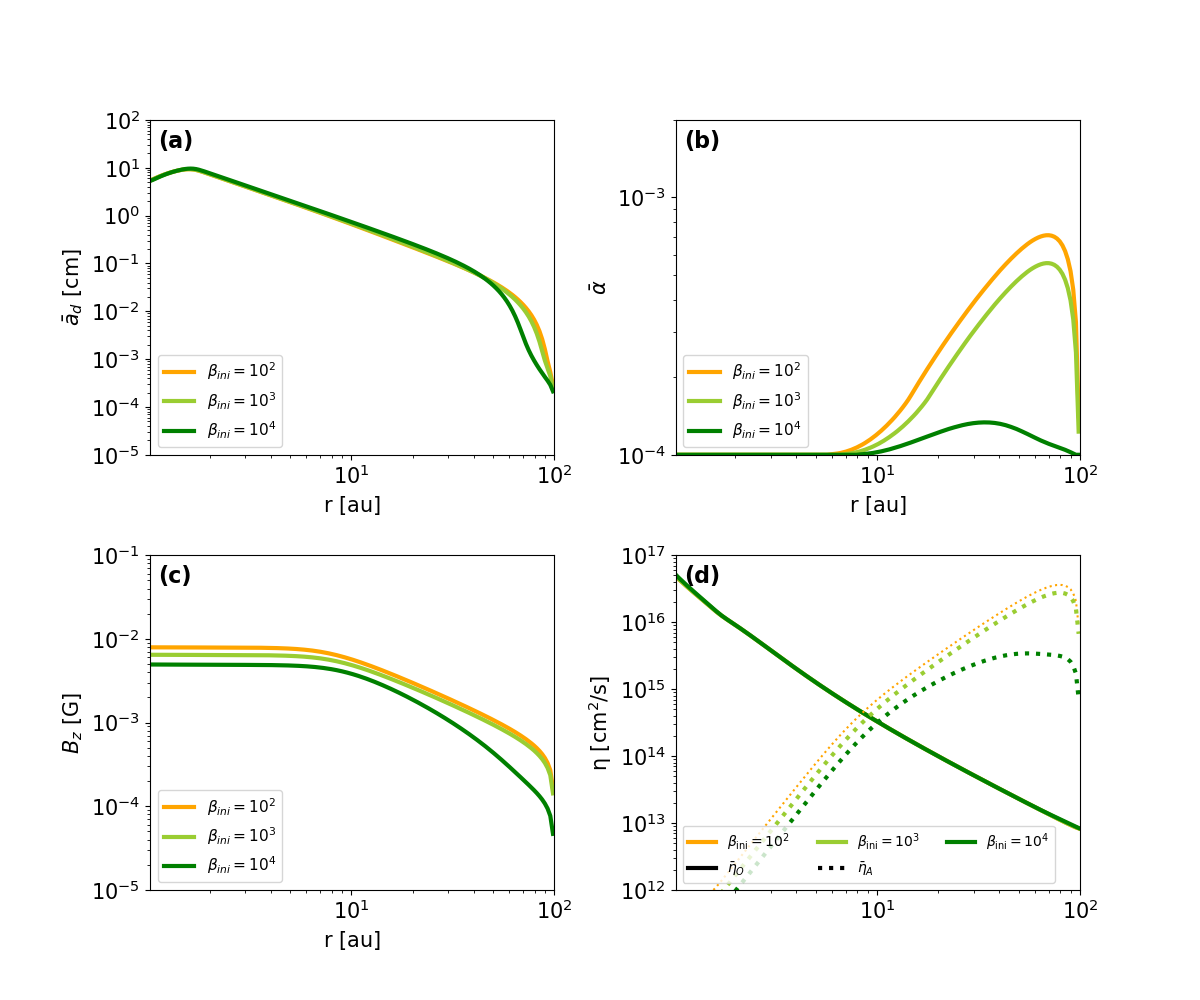}
  \caption{
  Radial distributions of 
(a) the mean dust size $\bar{a}_d$, 
(b) the vertically averaged turbulent viscosity $\bar{\alpha}$, 
(c) the vertical magnetic field strength $B_z$, and 
(d) the vertically averaged magnetic diffusivities 
$\bar{\eta}_O$ (solid lines) and $\bar{\eta}_A$ (dotted lines) 
at $t = 3.0\times10^5\,\mathrm{yr}$ for models with 
different initial plasma beta values: 
$\beta_{\mathrm{ini}} = 10^2$ (orange), 
$10^3$ (yellow-green), and $10^4$ (green).
  }
  \label{fig:comp_B}
\end{figure*}

In this subsection, we compare models with different 
initial magnetic field strengths (b2-frg30, b3-frg30, 
and b4-frg30) to examine how the initial magnetic field 
strength affects the dust size distribution, turbulent 
viscosity, and magnetic diffusivities.
Figure~\ref{fig:comp_B} presents the radial distributions of 
the mean dust size (Figure~\ref{fig:comp_B}a), the vertically 
averaged turbulent viscosity (Figure~\ref{fig:comp_B}b), 
the magnetic field strength (Figure~\ref{fig:comp_B}c), 
and the magnetic diffusivities (Figure~\ref{fig:comp_B}d) 
at $t = 3.0\times10^5\,\mathrm{yr}$ for the three models. 
We show that these quantities are closely coupled through 
the mutual interactions among magnetic-field evolution, 
MRI-driven turbulence, and dust evolution.

The mean dust size (Figure~\ref{fig:comp_B}a) is essentially 
independent of $\beta_{\mathrm{ini}}$ for 
$r \lesssim 60\,\mathrm{au}$. At larger radii, however, 
model b4-frg30 yields systematically smaller dust grains than 
models b2-frg30 and b3-frg30. This difference is associated with 
the turbulent viscosity (Figure~\ref{fig:comp_B}b), which is 
substantially lower in model b4-frg30 in the outer disk ($r\gtrsim10 \mathrm{au}$), 
reaching a peak value nearly an order of magnitude smaller than 
those in models b2-frg30 and b3-frg30. The weaker turbulence 
reduces the relative velocities between dust grains and thereby 
slows dust growth in the outer disk.

The magnetic field strength (Figure~\ref{fig:comp_B}c) is broadly 
similar among the three models in the inner disk. In the outer disk, 
however, model b4-frg30 exhibits systematically lower magnetic field 
strengths than models b2-frg30 and b3-frg30, consistent with its 
weaker initial magnetic field. The stronger magnetic fields in models 
b2-frg30 and b3-frg30 sustain stronger MRI-driven turbulence at 
$t = 3.0\times10^5\,\mathrm{yr}$, whereas the weaker magnetic field 
in model b4-frg30 results in weaker MRI-driven turbulence. 
Consequently, dust growth proceeds more rapidly in models b2-frg30 
and b3-frg30, producing larger grains in the outer disk than in 
model b4-frg30.

The two magnetic diffusivities (Figure~\ref{fig:comp_B}d) show 
different dependences on $\beta_{\mathrm{ini}}$. 
The Ohmic diffusivity $\bar{\eta}_O$ (solid lines) shows almost no 
dependence on $\beta_{\mathrm{ini}}$, with the curves for all three 
models nearly overlapping across the entire disk. 
By contrast, the ambipolar diffusivity $\bar{\eta}_A$ (dotted lines) 
depends appreciably on $\beta_{\mathrm{ini}}$ in the outer disk. 
For $r \gtrsim 40\,\mathrm{au}$, $\bar{\eta}_A$ reaches values of 
order $10^{16}\,\mathrm{cm^2\,s^{-1}}$ in models b2-frg30 and 
b3-frg30, whereas it remains roughly an order of magnitude lower, 
of order $10^{15}\,\mathrm{cm^2\,s^{-1}}$, in model b4-frg30. 
This variation in $\bar{\eta}_A$ among the three models is consistent 
with the $\beta_{\mathrm{ini}}$ dependence of the magnetic field 
strength (Figure~\ref{fig:comp_B}c), because the ambipolar diffusivity, 
unlike the Ohmic diffusivity, depends directly on the magnetic field 
strength.

Our results therefore demonstrate that the initial magnetic field 
strength can affect dust growth through its influence on MRI-driven 
turbulence, particularly in the outer disk. 
A related connection between dust growth and MRI-driven turbulence 
has been reported by \citet{Delage2023}, who found that MRI-driven 
turbulence develops preferentially in regions where dust growth 
proceeds efficiently. 
Their study, however, did not solve for the time evolution of the 
magnetic field or specify an initial field strength. 
Instead, they assumed a seed magnetic field sustaining MRI-driven 
accretion and derived the magnetic field strength required to sustain 
an assumed accretion rate. 
In their framework, the magnetic field strength is therefore 
determined by the assumed accretion rate and cannot be varied 
independently. 
In contrast, we solve the induction equation and follow the evolution 
of the magnetic field self-consistently together with MRI-driven 
turbulence and dust growth, without assuming a fixed accretion rate.
Our results show that variations in the initial magnetic field 
strength can lead to differences in the resulting dust size 
distribution. This effect is difficult to capture within a 
fixed-accretion-rate framework such as that of \citet{Delage2023}.

We note, however, that the dependence of dust growth on the initial 
magnetic field strength is largely confined to the outer disk 
($r \gtrsim 60\,\mathrm{au}$), where differences in MRI-driven 
turbulence affect grain growth. 
Over most of the disk, the dust size is primarily regulated by radial 
drift and fragmentation and therefore shows little dependence on 
$\beta_{\mathrm{ini}}$ (Figure~\ref{fig:comp_B}a). 
Magnetic field strength therefore plays only a limited 
role in determining the overall dust size distribution.

\begin{figure*}[t]
  \centering
  \includegraphics[width=0.7\textwidth]
  {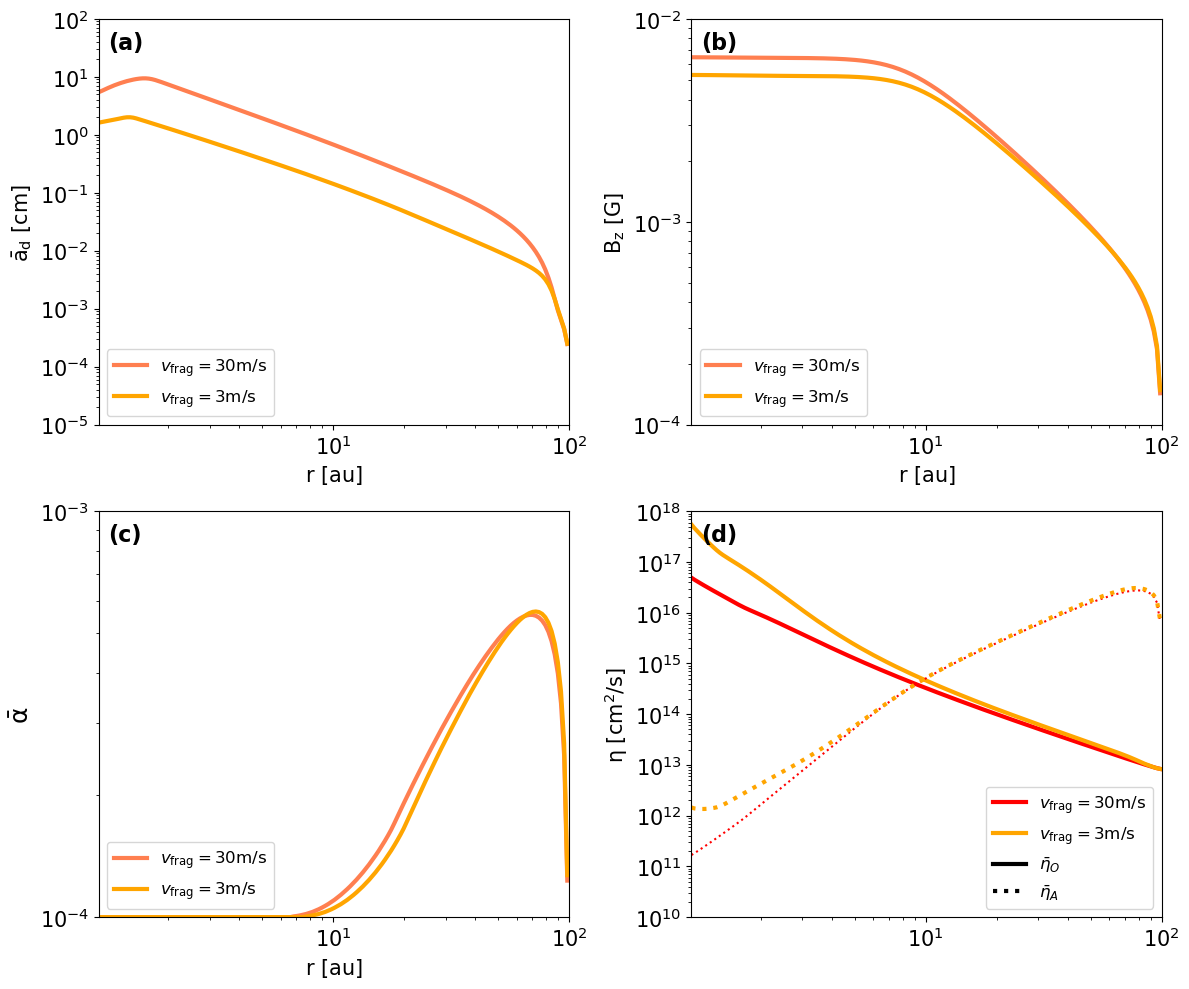}
  \caption{
  Radial distributions of 
(a) the mean dust size $\bar{a}_d$, 
(b) the vertical magnetic field strength $B_z$, 
(c) the vertically averaged turbulent viscosity $\bar{\alpha}$, 
and (d) the vertically averaged magnetic diffusivities 
$\bar{\eta}_O$ (solid lines) and $\bar{\eta}_A$ (dotted lines) 
at $t = 3.0\times10^5\,\mathrm{yr}$ for models with different 
fragmentation velocities: model b3-frg30 
($v_{\mathrm{frag}} = 30\,\mathrm{m\,s^{-1}}$, red) 
and model b3-frg3 
($v_{\mathrm{frag}} = 3\,\mathrm{m\,s^{-1}}$, orange).
  }
  \label{fig:comp_frg}
\end{figure*}

\subsection{Dependence on Fragmentation Velocity}
\label{subsec:dependence_vfrag}
We compare models b3-frg30 and b3-frg3 to examine how the 
fragmentation velocity affects the dust size distribution, magnetic 
field strength, turbulent viscosity, and magnetic diffusivities. 
The two models have the same initial plasma beta 
($\beta_{\mathrm{ini}} = 10^3$) but different fragmentation 
velocities ($v_{\mathrm{frag}} = 30\,\mathrm{m\,s^{-1}}$ and 
$3\,\mathrm{m\,s^{-1}}$, respectively). 
Figure~\ref{fig:comp_frg} shows the radial distributions of 
(a) the mean dust size, (b) the magnetic field strength, 
(c) the turbulent viscosity, and (d) the magnetic diffusivities 
at $t = 3.0\times10^5\,\mathrm{yr}$.

The mean dust size (Figure~\ref{fig:comp_frg}a) exhibits a clear 
dependence on $v_{\mathrm{frag}}$ over most of the disk. 
In model b3-frg30, $\bar{a}_d$ reaches a maximum of 
$\simeq 10\,\mathrm{cm}$ at $r \simeq 2\,\mathrm{au}$, whereas 
in model b3-frg3, it reaches a substantially smaller maximum of 
$\simeq 2\,\mathrm{cm}$ at a comparable radius. 
At larger radii, the mean dust size remains systematically smaller 
in model b3-frg3 than in model b3-frg30, with the difference 
diminishing only near the outer edge of the disk 
($r \gtrsim 90\,\mathrm{au}$). 
This difference arises because the lower fragmentation velocity in 
model b3-frg3 leads to more efficient fragmentation, which limits 
dust growth to smaller sizes.

Because the lower fragmentation velocity in model b3-frg3 leads to 
more efficient fragmentation, small grains are replenished more 
efficiently, resulting in a higher abundance of small grains in the 
inner disk ($r \lesssim 10\,\mathrm{au}$). 
The enhanced abundance of small grains reduces the ionization degree 
and thereby increases the magnetic diffusivities. 
Consequently, magnetic diffusion is enhanced in model b3-frg3, 
resulting in a slightly weaker magnetic field in the inner disk than 
in model b3-frg30 (Figure~\ref{fig:comp_frg}b). 
The difference in magnetic field strength is at most a factor of 2.

The higher abundance of small grains in model b3-frg3 reduces the 
ionization degree by absorbing charged particles, leading to enhanced 
magnetic diffusivities in the inner disk (Figure~\ref{fig:comp_frg}d). 
The effect is particularly pronounced for Ohmic resistivity: 
$\bar{\eta}_O$ in model b3-frg3 exceeds that in model b3-frg30 by more than an order of magnitude at $r \simeq 1\,\mathrm{au}$. 
This difference decreases with increasing radius and becomes negligible beyond $r \simeq 20\,\mathrm{au}$. 
Similar to $\bar{\eta}_O$, the ambipolar diffusivity $\bar{\eta}_A$ is also higher for the lower fragmentation velocity in model b3-frg3 in the region $r \lesssim 10\,\mathrm{au}$, because the higher abundance of
small grains reduces the ionization degree. Beyond $r \gtrsim
10\,\mathrm{au}$, however, the difference in $\bar{\eta}_A$ between the two models becomes negligible. The magnetic field strength therefore remains comparable between the two models across most of the outer disk (Figure~\ref{fig:comp_frg}b).

The turbulent viscosity (Figure~\ref{fig:comp_frg}c), in contrast, shows little variation between the two models throughout the disk. 
This is because MRI activity is governed primarily by the magnetic field strength and ionization degree, rather than directly by the fragmentation velocity.

In the outermost region of the disk ($r \gtrsim 80\,\mathrm{au}$), 
the physical quantities shown in Figure~\ref{fig:comp_frg} show 
relatively little dependence on $v_{\mathrm{frag}}$. 
This similarity arises because the collision velocities between dust 
grains remain below $v_{\mathrm{frag}}$ in both models in this region, 
so that fragmentation has little effect on dust growth. 
Consequently, the difference in mean dust size between the two models 
decreases toward the outer edge and becomes very small at 
$r \gtrsim 90\,\mathrm{au}$. 
The magnetic field strength, turbulent viscosity, and magnetic diffusivities also show little difference between the two models in this outermost region.

\begin{figure*}[t]
  \centering
  \includegraphics[width=0.7\textwidth]
  {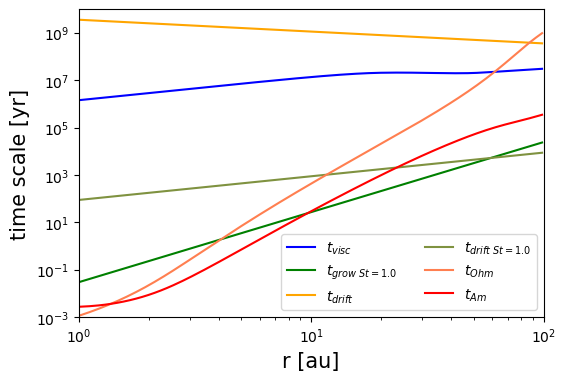}
  \caption{
  Radial distributions of relevant timescales for the fiducial 
model (b3-frg30), with $a_0 = 1.0\times10^{-5}\,\mathrm{cm}$ and 
$\beta_{\mathrm{ini}} = 10^3$. 
The viscous diffusion timescale $t_{\mathrm{vis}}$ (blue), 
radial drift timescale $t_{\mathrm{drift}}$ (orange), 
Ohmic diffusion timescale $t_{\mathrm{Ohm}}$ (pink), and 
ambipolar diffusion timescale $t_{\mathrm{Am}}$ (red) are 
evaluated at $t=0$ using the turbulent viscosity 
$\bar{\alpha}$ and magnetic diffusivities 
$\bar{\eta}_O$ and $\bar{\eta}_A$. 
The dust growth timescale $t_{\mathrm{grow,St=1}}$ (lime green) 
is the time required for grains initially of size $a_0$ to grow 
to $\mathrm{St}=1$ under the single-size approximation described 
in Section~\ref{subsec:time_scale}. 
The corresponding drift timescale 
$t_{\mathrm{drift,St=1}}$ (green) is the radial drift timescale 
of grains with $\mathrm{St}=1$ and does not include the time 
required for the grains to grow to this size.
    }
  \label{fig:timescale}
\end{figure*}

\subsection{Timescales}
\label{subsec:time_scale}
In this subsection, we compare the characteristic timescales of dust 
growth, radial drift, viscous evolution, and magnetic diffusion at 
each radius. This comparison serves two purposes. First, we assess 
the validity of neglecting the global evolution of the gas in our 
model by comparing the viscous timescale with the characteristic 
timescales of dust and magnetic-field evolution. Second, we compare 
the timescales of dust growth, radial drift, and magnetic diffusion 
to identify which processes act first during the early evolution and 
thereby provide a physical interpretation of the results presented 
above. Although the timescales themselves evolve as the dust and 
magnetic field evolve, the comparison is also useful for understanding 
the qualitative interplay among dust growth, radial drift, and magnetic diffusion during the subsequent evolution.

Unless otherwise noted, the turbulent viscosity $\bar{\alpha}$ and 
the magnetic diffusivities $\bar{\eta}_O$ and $\bar{\eta}_A$ used to 
evaluate the timescales below are evaluated at their initial values 
($t=0$). The timescale comparison therefore characterizes the 
relative importance of the relevant processes during the early 
evolution and does not necessarily represent their ordering 
throughout the subsequent evolution. 
Figure~\ref{fig:timescale} shows the radial distributions of these 
timescales for the fiducial model b3-frg30, using 
$a_0 = 1.0\times10^{-5}\,\mathrm{cm}$ and 
$\beta_{\mathrm{ini}} = 10^3$.

The viscous diffusion timescale and the radial drift timescale of 
monomer-sized grains are given by
\begin{align}
t_{\mathrm{vis}} &= \frac{r^{2}}{\nu_t}, \\
t_{\mathrm{drift}} &= \frac{r}{v_{r}(m_0)},
\end{align}
where $m_0$ is the dust mass corresponding to the monomer size 
$a_0 = 1.0\times10^{-5}\,\mathrm{cm}$.

Following \citet{Sato2016}, the dust growth timescale under the 
single-size approximation, in which the relative velocity is evaluated 
between equal-sized particles, is given by
\begin{align}
t_{\mathrm{grow}}(m_0)
= \left(\frac{1}{a_d}\frac{\mathrm{d}a_d}{\mathrm{d}t}
\right)^{-1}
= \frac{3m_0 H_d(m_0)}{2\sqrt{\pi}\,\Delta v_{0}\,
\Sigma_d},
\end{align}
where $\Delta v_0$ is the relative velocity between two grains of 
mass $m_0$. Dust grains drift most efficiently when 
$\mathrm{St}\simeq1$. Assuming the Epstein regime, the corresponding 
dust size is given by
\begin{align}
a_{\mathrm{St=1}} = \frac{2\Sigma_g}{\pi\rho_{\mathrm{int}}}.
\end{align}
Assuming that $t_{\mathrm{grow}}$ remains approximately constant 
during growth, we estimate the time required for particles to grow 
from $a_0$ to $a_{\mathrm{St=1}}$ as
\begin{align}
t_{\mathrm{grow,St=1}}
= t_{\mathrm{grow}}(m_0)\,
\log\!\left(\frac{a_{\mathrm{St=1}}}{a_0}\right).
\end{align}
This expression provides an order-of-magnitude estimate of the time 
required for particles to reach $\mathrm{St}=1$. 
The corresponding radial drift timescale for grains with $\mathrm{St}=1$ is
\begin{align}
t_{\mathrm{drift,St=1}} =
\frac{r}{v_r(a_{\mathrm{St=1}})}.
\end{align}

The magnetic diffusion timescales due to Ohmic resistivity and 
ambipolar diffusion are given by
\begin{align}
t_{\mathrm{Ohm}} &= \frac{r^{2}}{\bar\eta_O},\\
t_{\mathrm{Am}}  &= \frac{r^{2}}{\bar\eta_A}.
\end{align}
These diffusivities are evaluated for the initial dust population 
consisting of monomer-sized grains 
($a_0 = 1.0\times10^{-5}\,\mathrm{cm}$). Under these initial 
conditions, ambipolar diffusion is faster than Ohmic diffusion over 
most of the disk, i.e., $t_{\mathrm{Am}} < t_{\mathrm{Ohm}}$. 
We note that this ordering differs from that obtained for the evolved 
dust size distribution discussed in 
Section~\ref{subsec:fiducial model}, where Ohmic resistivity dominates 
over a wider region of the inner disk.

\subsubsection{Justification of model assumptions}
In Section~\ref{subsec:evolution of magnetic field}, we assumed that 
the global radial evolution of the gas can be ignored, both in the 
induction equation for the magnetic field and in the equations 
governing dust evolution. Here, we use the timescale comparison shown 
in Figure~\ref{fig:timescale} to assess the validity of this assumption.

Over most of the disk, the viscous diffusion timescale 
$t_{\mathrm{vis}}$ is much longer than the magnetic diffusion 
timescales ($t_{\mathrm{Ohm}}$ and $t_{\mathrm{Am}}$) and the dust 
growth timescale $t_{\mathrm{grow,St=1}}$. Thus, the global gas 
distribution evolves only slowly through viscous evolution while 
dust growth and magnetic diffusion proceed, supporting our 
approximation of neglecting the global radial evolution of the gas.

At $r \gtrsim 60$--$70\,\mathrm{au}$, $t_{\mathrm{Ohm}}$ exceeds 
$t_{\mathrm{vis}}$ (Figure~\ref{fig:timescale}). However, 
$t_{\mathrm{Am}}$, which provides the shorter magnetic diffusion 
timescale in this region, remains much shorter than 
$t_{\mathrm{vis}}$. Thus, magnetic diffusion can still proceed much 
faster than the global viscous evolution of the gas in the outer disk.

This argument may not apply to magnetic-flux transport in the disk 
surface layers. In our model, the gas density distribution is fixed 
and radial gas motion is neglected. However, \citet{Guilet2012} 
showed that magnetic flux can be rapidly advected by gas in the 
surface layers of protoplanetary disks. Such surface-layer advection, 
if present, could affect the radial evolution of $B_z$ even when the 
global viscous evolution of the disk is slow. Since gas dynamics is 
not solved in our model, this effect is not included in the present 
calculations. A quantitative assessment of this effect is left for 
future work.

\subsubsection{Coupling between dust growth and magnetic diffusion}
In this subsubsection, we compare the dust growth timescale
$t_{\mathrm{grow,St=1}}$ (the time required for dust to grow to
$\mathrm{St}=1$) with the Ohmic diffusion timescale $t_{\mathrm{Ohm}}$
and the ambipolar diffusion timescale $t_{\mathrm{Am}}$ to examine
how dust growth can affect the magnetic field evolution at different radii.

In the region $r \lesssim 4\,\mathrm{au}$, both magnetic diffusion
processes proceed faster than dust growth,
\begin{align}
t_{\mathrm{Am}} \lesssim t_{\mathrm{Ohm}} \lesssim t_{\mathrm{grow,St=1}}.
\end{align}
In this region, magnetic diffusion proceeds before dust growth can significantly alter the ionization degree. Therefore, changes in the magnetic diffusivities caused by dust growth have little effect on the magnetic field evolution.

In the region $4\,\mathrm{au} \lesssim r \lesssim 10\,\mathrm{au}$, the ambipolar diffusion timescale is comparable to or shorter than the dust growth timescale, while the Ohmic diffusion timescale becomes substantially longer than $t_{\mathrm{grow,St=1}}$,
\begin{align}
t_{\mathrm{Am}} \lesssim t_{\mathrm{grow,St=1}}
\lesssim t_{\mathrm{Ohm}}.
\end{align}
As dust grows, the abundance of small grains decreases, which increases the ionization degree and thereby reduces the magnetic diffusivities. Because ambipolar diffusion proceeds on a timescale comparable to or shorter than dust growth ($t_{\mathrm{Am}} \lesssim t_{\mathrm{grow,St=1}}$), dust growth has little effect on the magnetic field evolution through changes in the ambipolar diffusivity. By contrast, because dust growth proceeds faster than Ohmic diffusion ($t_{\mathrm{grow,St=1}} \lesssim t_{\mathrm{Ohm}}$), dust growth can reduce $\bar\eta_O$ before Ohmic diffusion proceeds significantly. Since Ohmic resistivity suppresses MRI turbulence, the resulting decrease in $\bar\eta_O$ can allow the MRI to become active. This can explain the results shown in Figures~\ref{fig:fiducial_evol}, \ref{fig:beta2_evol}, and \ref{fig:beta4_evol}. In each of these models, regions that are MRI inactive at $t=0$ become turbulent by $t=1.0\times10^4\,\mathrm{yr}$.

In the region $r \gtrsim 10\,\mathrm{au}$, dust growth proceeds faster than magnetic diffusion,
\begin{align}
t_{\mathrm{grow,St=1}} \lesssim t_{\mathrm{Am}} < t_{\mathrm{Ohm}}.
\end{align}
Dust growth in this region can significantly alter the ionization
degree, and hence the magnetic diffusivities, before either magnetic
diffusion process proceeds significantly. 
Because $t_{\mathrm{Am}} < t_{\mathrm{Ohm}}$ in this region, ambipolar
diffusion, rather than Ohmic resistivity, governs the magnetic field
evolution. 
Thus, dust growth can affect MRI activity in this
region mainly through changes in the ambipolar diffusivity
$\bar\eta_A$, rather than in the Ohmic resistivity $\bar\eta_O$.

\subsubsection{Coupling between radial drift and magnetic diffusion}
In this subsubsection, we compare the drift timescale of dust grains that have grown to $\mathrm{St}=1$, $t_{\mathrm{drift,St=1}}$, with the Ohmic diffusion timescale $t_{\mathrm{Ohm}}$ and the ambipolar diffusion timescale $t_{\mathrm{Am}}$ to assess how the radial drift of such particles may affect the magnetic field evolution.

In the region $r \lesssim 10\,\mathrm{au}$, magnetic diffusion
proceeds faster than radial drift,
\begin{align}
t_{\mathrm{Am}} \lesssim t_{\mathrm{Ohm}} \lesssim 
t_{\mathrm{drift,St=1}}.
\end{align}
In this region, magnetic diffusion proceeds before radial drift can
significantly alter the ionization degree. Therefore, changes in the
magnetic diffusivities caused by radial drift have little effect on
the magnetic field evolution.

In the region $10\,\mathrm{au} \lesssim r \lesssim 20\,\mathrm{au}$,
$t_{\mathrm{drift,St=1}}$ decreases relative to the magnetic diffusion
timescales, so that radial drift proceeds faster than Ohmic diffusion
but slower than ambipolar diffusion,
\begin{align}
t_{\mathrm{Am}} \lesssim t_{\mathrm{drift,St=1}}
\lesssim t_{\mathrm{Ohm}}.
\end{align}
Because $t_{\mathrm{drift,St=1}} \lesssim t_{\mathrm{Ohm}}$, radial
drift removes large dust grains and thereby alters the ionization
degree before Ohmic diffusion proceeds significantly. 
Thus, radial drift can affect the magnetic field evolution through changes in Ohmic resistivity. 
By contrast, because
$t_{\mathrm{Am}} \lesssim t_{\mathrm{drift,St=1}}$, ambipolar
diffusion proceeds on a timescale comparable to or shorter than
radial drift. Therefore, changes in the ambipolar diffusivity caused
by radial drift have little effect on the magnetic field evolution
in this region.

In the region $r \gtrsim 20\,\mathrm{au}$, radial drift proceeds
faster than both magnetic diffusion processes,
\begin{align}
t_{\mathrm{drift,St=1}} \lesssim t_{\mathrm{Am}} < t_{\mathrm{Ohm}}.
\end{align}
Dust grains reaching $\mathrm{St}=1$ therefore drift inward before either
magnetic diffusion process proceeds significantly, reducing the dust
abundance in the outer region of the disk. 
This drift-induced decrease in dust abundance can alter the ionization degree, and hence the magnetic diffusivities, before magnetic diffusion proceeds significantly. 
Because $t_{\mathrm{Am}} < t_{\mathrm{Ohm}}$ in this region, ambipolar diffusion, rather than Ohmic resistivity, governs the magnetic field evolution. 
Thus, radial drift can affect MRI
activity in this region mainly through changes in the ambipolar
diffusivity.

\subsection{Ambipolar Drift Velocity} \label{subsec:effect of ambipolar_drift}
\begin{figure*}[t]
  \centering
  \includegraphics[width=\textwidth]{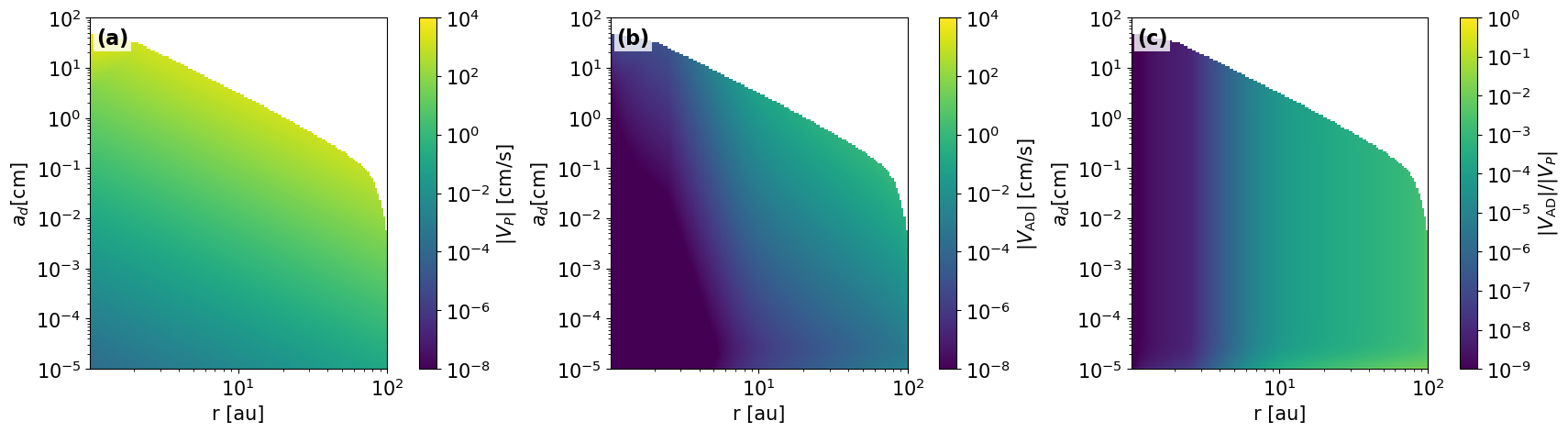}
  \caption{
   Drift velocities and their ratio as functions of grain radius $a_\mathrm{d}$ and orbital radius $r$ in the disk midplane of the fiducial model (b3-frg30) at $t = 3.0 \times 10^5$\,yr. Panels (a), (b), and (c) show the pressure-driven drift velocity $|V_\mathrm{P}|$, the ambipolar drift velocity $|V_\mathrm{AD}|$, and the ratio $|V_\mathrm{AD}|/|V_\mathrm{P}|$, respectively. 
  }
  \label{fig:Ambipolar_drift}
\end{figure*}

In this subsection, we evaluate the effect of ambipolar drift on dust growth.
For this purpose, we compare the pressure-driven drift velocity $V_\mathrm{P}$ and the ambipolar drift velocity $V_\mathrm{AD}$ of dust grains in the disk midplane at $t = 3.0 \times 10^5$\,yr for the fiducial model (b3-frg30).
The magnitudes of $V_\mathrm{P}$ and $V_\mathrm{AD}$ are given by
\begin{align}
|V_\mathrm{P}| &= \sqrt{v_r^2 + v_\phi^2}, \\
|V_\mathrm{AD}| &= \sqrt{v_{r,\mathrm{AD}}^2 + v_{\phi,\mathrm{AD}}^2},
\end{align}
where $v_r$ and $v_\phi$ are given by Equations~(\ref{eq:vr}) and (\ref{eq:vphi}), and $v_{r,\mathrm{AD}}$ and $v_{\phi,\mathrm{AD}}$ are the radial and azimuthal components of the ambipolar drift velocity, respectively.
We calculate $v_{r,\mathrm{AD}}$ and $v_{\phi,\mathrm{AD}}$ from Equations~(\ref{eq:vr_AD}) and (\ref{eq:vphi_AD}), which we derive following the method of \citet{Silsbee2020} (Appendix~\ref{app:calculation of ambipolar drift}).

Figure~\ref{fig:Ambipolar_drift} shows (a) $|V_\mathrm{P}|$, (b) $|V_\mathrm{AD}|$, and (c) their ratio $|V_\mathrm{AD}|/|V_\mathrm{P}|$ as functions of orbital radius and grain size.
In panel (a), $|V_\mathrm{P}|$ reaches $10^2$\,cm\,s$^{-1}$ or more for some grains with $a_\mathrm{d} \gtrsim 1$\,cm.
In panel (b), $|V_\mathrm{AD}|$ generally increases with orbital radius at a fixed grain size within the plotted region. At a given orbital radius in the outer disk, it tends to be larger for larger grains.
Panel (c) shows that $|V_\mathrm{AD}|$ is much smaller than $|V_\mathrm{P}|$ throughout the plotted region. 
The small ambipolar drift velocities suggest that ambipolar drift is unlikely to substantially increase the collision frequency of dust grains under the conditions examined here.

\citet{Silsbee2020} and \citet{Guillet2020} calculated the drift of charged dust grains in molecular cloud cores, including the Lorentz force. They found that grains smaller than about $0.1\,\mu\mathrm{m}$ are coupled to the magnetic field and are rapidly depleted as they are incorporated into larger grains through coagulation. Our model has a minimum grain radius of $0.1\,\mu\mathrm{m}$ and thus excludes grains smaller than this value. In addition, the gas density is higher in the disk midplane than in molecular cloud cores, which strengthens gas drag. 
The small magnitudes of the ambipolar drift velocities shown in Figure~\ref{fig:Ambipolar_drift} suggest that this mechanism plays only a minor role in grain coagulation under the conditions examined here.

\subsection{Electrostatic Barrier} \label{subsec:effect of electrostatic barrier}

\begin{figure*}[t]
  \centering
  \includegraphics[width=\textwidth]{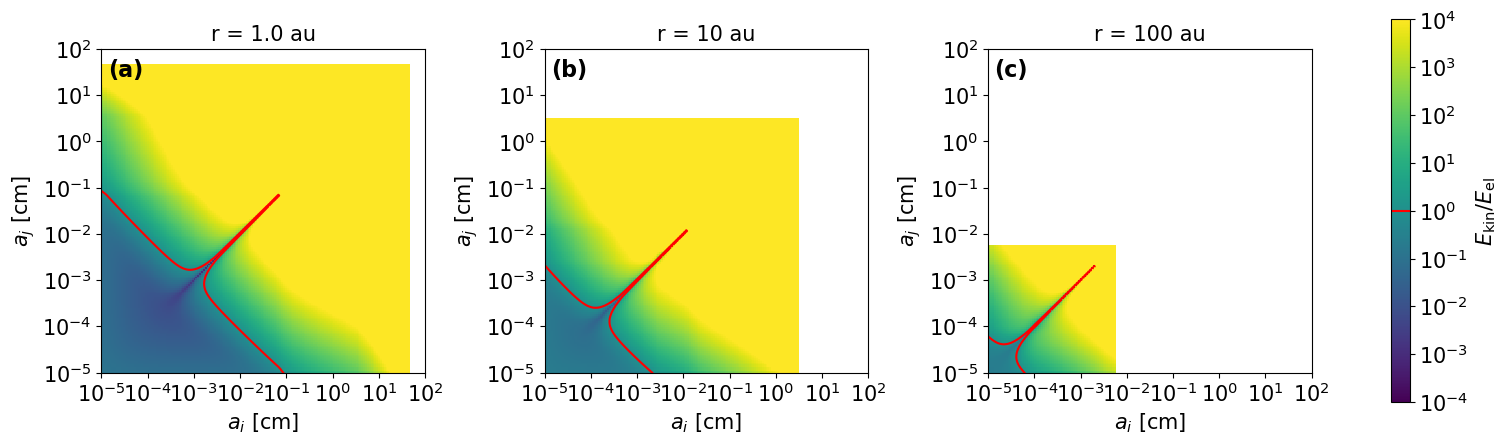}
\caption{
Ratio of the collision kinetic energy to the electrostatic energy, $E_\mathrm{kin}/E_\mathrm{el}$, as a function of the grain radii $a_i$ and $a_j$ in the disk midplane of the fiducial model (b3-frg30) at $t = 3.0 \times 10^5$\,yr. Panels (a), (b), and (c) correspond to orbital radii of 1.0, 10, and 100\,au, respectively. The red contours mark $E_\mathrm{kin} = E_\mathrm{el}$.
}
  \label{fig:electrostatic_barrier}
\end{figure*}

To assess the effect of the electrostatic barrier on dust growth, we compare the electrostatic energy $E_\mathrm{el}$ with the collision kinetic energy $E_\mathrm{kin}$ in the midplane of the fiducial model (b3-frg30) at $t = 3.0 \times 10^5$\,yr. 
Following \citet{Okuzumi2009}, two colliding dust grains cannot come into contact when $E_\mathrm{el} > E_\mathrm{kin}$.

For two dust grains with masses $m_i$ and $m_j$ and radii $a_i$ and $a_j$, the collision kinetic energy and electrostatic energy are
\begin{align}
E_\mathrm{kin} &= \frac{1}{2} \frac{m_i m_j}{m_i + m_j} (\Delta V_{i,j})^2, \\
E_\mathrm{el} &= \frac{Z_i Z_j e^2}{a_i + a_j},
\end{align}
where $\Delta V_{i,j}$ is the total relative velocity and $e$ is the elementary charge. We obtain the mean charge numbers $Z_i$ and $Z_j$ for the respective grain sizes from the chemical equilibrium calculation described in Section~\ref{subsec: calculation of non-ideal effect}.
The total relative velocity satisfies
\begin{equation}
(\Delta V_{i,j})^2 = (\Delta V^\mathrm{B}{i,j})^2 + (\Delta V^\mathrm{T}{i,j})^2 + (\Delta V^\mathrm{D}_{i,j})^2.
\end{equation}

Figure~\ref{fig:electrostatic_barrier} shows $E_\mathrm{kin}/E_\mathrm{el}$ as a function of the two grain radii at orbital radii of (a) 1.0\,au, (b) 10\,au, and (c) 100\,au. 
The red contours mark $E_\mathrm{kin} = E_\mathrm{el}$. 
At all three orbital radii, $E_\mathrm{kin}/E_\mathrm{el} < 1$ for collisions between small grains, indicating that the collision energy is insufficient to overcome the electrostatic barrier. For pairs of nearly equal size, this condition extends to larger grain radii. 
Such grains have similar drift velocities. Therefore, differential drift contributes little to their relative velocity and hence to their collision kinetic energy.

The grain sizes for which $E_\mathrm{el} > E_\mathrm{kin}$ vary with orbital radius (Figure~\ref{fig:electrostatic_barrier}). 
For pairs of nearly equal size, the upper grain radius for which this inequality holds decreases from about $1.0\,\mathrm{mm}$ at 1.0\,au to $0.2\,\mathrm{mm}$ at 10\,au and $50\,\mu\mathrm{m}$ at 100\,au. 
For pairs of unequal size, the inequality holds for $d_{i,j} \equiv a_i + a_j$ below approximately $1.0 \, \mathrm{m m}$, $30 \, \mathrm{\mu m}$, and $0.8\,\mu\mathrm{m}$ at 1.0, 10, and 100\,au, respectively. 
The ratio $E_\mathrm{kin}/E_\mathrm{el}$ is generally smaller in the inner disk than in the outer disk, indicating a stronger electrostatic barrier relative to the collision energy in the inner disk.

\citet{Akimkin2020} modeled the evolution of the dust size distribution while accounting for both grain charging and disk ionization. 
They found that the electrostatic barrier suppresses coagulation among grains with radii of $0.1$--$10\,\mu\mathrm{m}$, whereas these grains can still coagulate with grains larger than $100\,\mu\mathrm{m}$. 
This dependence on the sizes of the colliding grains is consistent with our results.

Our energy comparison suggests that the electrostatic barrier can hinder coagulation between small grains, whereas collisions with $E_\mathrm{kin} > E_\mathrm{el}$ can overcome the barrier. Grains may grow in the outer disk, where collisions can more readily overcome the electrostatic barrier. Once grown, these grains may drift inward and serve as seeds for coagulation with small grains \citep{Akimkin2020}. Over most of the disk, the maximum grain size in our simulations is set by radial drift (Figure~\ref{fig:dsize}) and exceeds the sizes at which the barrier is effective in our estimate. 
Future work should test whether grains can reach the drift-limited size when electrostatic repulsion is included self-consistently in the dust evolution calculations.

\section{Conclusion}\label{sec:conclusion}
We performed one-dimensional calculations of the 
coupled evolution of the dust size distribution 
and the magnetic field in a protoplanetary 
disk. Our model includes the feedback of dust 
growth on non-ideal MHD effects and MRI-driven 
turbulence. We treated the initial plasma beta 
$\beta_{\mathrm{ini}}$ and the fragmentation 
velocity $v_{\mathrm{frag}}$ as free parameters. 
The main results are summarized as follows.

We find that, for models b2-frg30, b3-frg30, and 
b4-frg30, the MRI-driven turbulent viscosity is 
larger for smaller $\beta_{\mathrm{ini}}$ in the 
outer region of the disk ($r \gtrsim 10\,\mathrm{au}$). Correspondingly, 
dust growth is more efficient for smaller 
$\beta_{\mathrm{ini}}$ in this region. Once grains 
grow sufficiently large, however, radial drift 
dominates their evolution regardless of the local 
turbulence level. Radial drift weakens the dependence of 
dust size on $\beta_{\mathrm{ini}}$. As a result, 
the dependence of dust size on the initial magnetic 
field strength is significant only in the outer 
region of the disk.

We also find that a smaller fragmentation velocity 
$v_{\mathrm{frag}}$ results in a smaller mean dust size 
over most of the disk. In model b3-frg3, the smaller 
$v_{\mathrm{frag}}$ leads to more efficient fragmentation 
and increases the abundance of small grains. Although this 
enhances the magnetic diffusivities, particularly in the 
inner region of the disk, the resulting differences in the 
magnetic field strength and MRI-driven turbulent viscosity 
remain relatively small. In the outermost region of the disk, 
the relative velocities between dust grains rarely exceed 
$v_{\mathrm{frag}}$, and thus the dust size and magnetic 
diffusivities show little dependence on $v_{\mathrm{frag}}$. 
Overall, the MRI-driven turbulent viscosity depends only weakly 
on the fragmentation velocity.

Although the feedback of dust size evolution on non-ideal MHD effects
is included in our model, the maximum dust size over most of the disk
is ultimately determined by radial drift rather than by MRI-driven
turbulent viscosity. The region in which MRI-driven turbulence can
promote dust growth is confined to the outer disk, where radial drift
is comparatively inefficient. Even within this region, an excessively
strong magnetic field can suppress MRI-driven turbulence itself, as
found for the strong-magnetic-field model
(b2-frg30; Section~\ref{subsec:strong_model}). The promotion of dust
growth by MRI-driven turbulence is therefore limited to the epochs and
radii at which the magnetic field strength remains within the range that
sustains turbulence. Because these conditions are met only over a
limited fraction of the disk, we conclude that, under the conditions
explored in this study, the magnetic field strength has only a minor
overall effect on dust growth, whereas radial drift remains the dominant
factor controlling the maximum dust size across most of the disk.
\begin{acknowledgments}
We thank the referee for very useful comments on this paper. This work was supported by JST SPRING, Japan Grant Number JPMJSP2136. The present study was supported by JSPS KAKENHI Grant (JP25K07369: M.N.M.). This work was supported by a NAOJ ALMA Scientific Research grant (No. 2022-22B). 
\end{acknowledgments}

\bibliography{shimokawa2026_1}{}
\bibliographystyle{aasjournalv7}

\twocolumngrid
\appendix

\section{Ambipolar Drift Velocity of Charged dust grains}\label{app:calculation of ambipolar drift}

Following the method of \citet{Silsbee2020}, we derive the radial and azimuthal components of the ambipolar drift velocity, $v_{r,\mathrm{AD}}$ and $v_{\phi,\mathrm{AD}}$, introduced in Section~\ref{subsec:effect of ambipolar_drift}. 
We assume an axisymmetric disk in which the gas rotates at the Keplerian velocity with zero radial velocity. 
We define these two components in a frame rotating at the local Keplerian angular velocity, so that they describe the dust motion relative to a circular Keplerian orbit.

We assume that only gravity, gas drag, and the Lorentz force act on the dust grains. The Lorentz force is exerted by the vertical magnetic field $B_z$, which is the only component of the magnetic field in our disk model (Section~\ref{subsec:evolution of magnetic field}).
Gravity is balanced by the centrifugal force on the reference circular Keplerian orbit. We linearize the equations of motion about this orbit.

We denote the stopping time of a dust grain by $t_\mathrm{stop}$ (Equation~(\ref{eq:stoppingtime})) and the radial drift velocity of the magnetic field relative to the neutral gas due to ambipolar diffusion by $w$. In this coordinate system, the magnetic field lines drift radially with velocity $w$. The Lorentz force depends on the grain velocity relative to the field lines, with radial and azimuthal components $v_{r,\mathrm{AD}} - w$ and $v_{\phi,\mathrm{AD}}$, respectively. The signed gyrofrequency of a dust grain is given by
\begin{equation}
\Omega_\mathrm{g} = \frac{Z e B_z}{m c},
\end{equation}
where $Z$ is the mean charge number for the grain size, $e$ is the elementary charge, $m$ is the grain mass, and $c$ is the speed of light. We obtain $Z$ from the chemical equilibrium calculation described in Section~\ref{subsec: calculation of non-ideal effect}. The linearized equations of motion for $v_{r,\mathrm{AD}}$ and $v_{\phi,\mathrm{AD}}$ are then given by
\begin{align}
\frac{dv_{r,\mathrm{AD}}}{dt} &= 2\Omega v_{\phi,\mathrm{AD}} - \frac{v_{r,\mathrm{AD}}}{t_\mathrm{stop}} + \Omega_\mathrm{g} v_{\phi,\mathrm{AD}}, \label{eq:eom_r} \\
\frac{dv_{\phi,\mathrm{AD}}}{dt} &= -\frac{1}{2}\Omega v_{r,\mathrm{AD}} - \frac{v_{\phi,\mathrm{AD}}}{t_\mathrm{stop}} - \Omega_\mathrm{g} \left(v_{r,\mathrm{AD}} - w\right). \label{eq:eom_phi}
\end{align}

When the magnetic field has only the vertical component $B_z$, the ambipolar diffusion term in the induction equation (Equation~(\ref{eq:induction})) gives
\begin{equation}
w = -\frac{\bar{\eta}_A}{B_z}\frac{\partial B_z}{\partial r}.
\end{equation}

We define $\Gamma_\mathrm{d} = \Omega_\mathrm{g} t_\mathrm{stop}$, whose magnitude measures the coupling of a dust grain to the magnetic field relative to its coupling to the gas \citep{Guillet2020}. Setting the time derivatives in Equations~(\ref{eq:eom_r}) and (\ref{eq:eom_phi}) to zero gives the steady-state solutions
\begin{align}
v_{r,\mathrm{AD}} &= \frac{\Gamma_\mathrm{d}\,(2\mathrm{St} + \Gamma_\mathrm{d})}{D}\, w, \label{eq:vr_AD} \\
v_{\phi,\mathrm{AD}} &= \frac{\Gamma_\mathrm{d}}{D}\, w, \label{eq:vphi_AD} \\
D &= 1 + \mathrm{St}^2 + \frac{5}{2}\mathrm{St}\,\Gamma_\mathrm{d} + \Gamma_\mathrm{d}^2. \label{eq:D_AD}
\end{align}
\end{document}